\documentclass[sigplan,nonacm]{acmart}

\usepackage{physics} %
\usepackage{soul} %
\usepackage{graphicx}
\usepackage{subcaption} %
\usepackage{float} %
\usepackage{placeins} %

\usepackage{textcomp}
\usepackage{xcolor}
\usepackage{tikz}
\usepackage{comment}
\usepackage{enumitem}

\usepackage{algorithm}
\usepackage{algpseudocode}

\usepackage{booktabs}

\newcommand{\fone}{15-to-1}
\newcommand{\name}{GP$\ket T$-S}
\newcommand{\tstate}{$\ket T$-state}
\newcommand{\tstates}{$\ket T$-states}

\newcommand{\terror}{$T$-error}

\newcommand{\cultconfig}
{
\ensuremath{d_{\mathrm{cult}}=5,\ d_{\mathrm{esc}}=15,\ p=10^{-3}}
}
\newcommand{\setp}{$\mathcal{P}$}
\newcommand{\extend}{\texttt{EXTEND}}
\newcommand{\merge}{\texttt{MERGE}}
\DeclareMathOperator{\pred}{pred}
\newcommand{\nondom}{\textproc{Nondom}}

\long\def\comment#1{}

\def\parah#1{\vspace*{0.0in} \noindent{\bf #1:}}

\begin{document}
\title{Gap-Priced $\ket T$-States: Dynamic $\ket T$ Post-Selection and Budget-Aware Factory Provisioning}

\author{Maxwell Poster}
\affiliation{%
  \institution{The University of Texas at Austin}
  \city{Austin}
  \state{Texas}
  \country{United States}
  }
\email{mposter@utexas.edu}

\author{Sayam Sethi}
\affiliation{%
  \institution{The University of Texas at Austin}
  \city{Austin}
  \state{Texas}
  \country{United States}
}

\author{Steven Hofmeyr}
\affiliation{%
  \institution{Lawrence Berkeley National Laboratory}
  \city{Berkeley}
  \state{California}
  \country{United States}
}

\author{Yuhao Liu}
\affiliation{%
 \institution{University of Pennsylvania}
 \city{Philadelphia}
 \state{Pennsylvania}
 \country{United States}
 }

\author{Jonathan Baker}
\affiliation{%
  \institution{The University of Texas at Austin}
  \city{Austin}
  \state{Texas}
  \country{United States}
}

\author{Costin Iancu}
\affiliation{%
  \institution{Lawrence Berkeley National Laboratory}
  \city{Berkeley}
  \state{California}
  \country{United States}
}

\renewcommand{\shortauthors}{Poster et al.}

\begin{abstract}
Magic \tstate{} preparation dominates the cost of fault-tolerant quantum computation on surface codes, yet cultivation and distillation are typically provisioned statically, leaving much of the program's error budget unused. We present Gap-Priced $\ket T$ States (\name{}), a lattice surgery scheduler and factory allocator that co-optimizes both methods under a total $T$-error budget. \name{} uses the complementary gap of each cultivated state as a fidelity estimate and dynamically adjusts the cultivation gap cutoff, converting budget slack into speed by giving the scheduler control over the cultivation post-selection. \name{} also characterizes distillation factories by size, $\ket T$-state production rate, and budget slack, and introduces cultivating factories that embed cultivation within routing channels, achieving fidelities otherwise requiring two-level distillation at a single-level footprint. \name{} organizes these characterizations into a factory catalog, and performs distillation factory allocation over this catalog in tandem with the gap cutoff using dynamic-programming to achieve one of i) minimum spacetime volume, ii) minimum execution time, or iii) minimum space overhead, all within a user-provided \terror{} budget. Across sixteen benchmarks under high and low \terror{} budgets of $10^{-3}$ and $10^{-5}$ respectively, \name{} consistently outperforms cultivation-only and single-factory baselines without exceeding the budget, achieving a 5.5$\times$ geometric mean speedup in logical cycle count and a 3.4$\times$ geometric mean reduction in space-time volume.  
\end{abstract}

\maketitle

\begin{figure}[tb]
    \centering
    \begin{subfigure}[t]{\columnwidth}
        \includegraphics[width=\linewidth,clip]{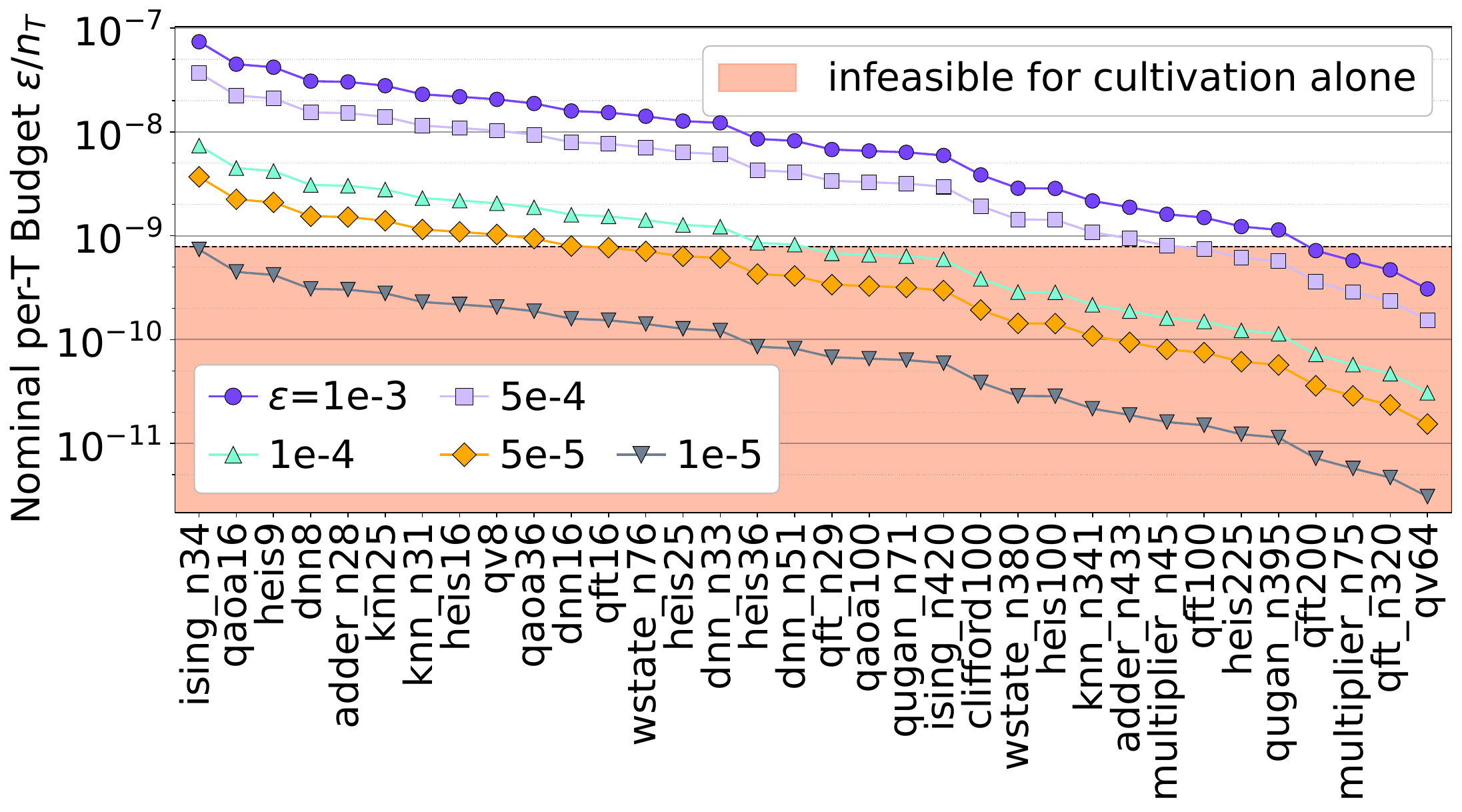}
        \vspace{-18pt}
        \caption{Cultivation feasibility
            \label{fig:cultfeas}}
    \end{subfigure}
    \begin{subfigure}[t]{\columnwidth}
        \includegraphics[width=\linewidth,clip]{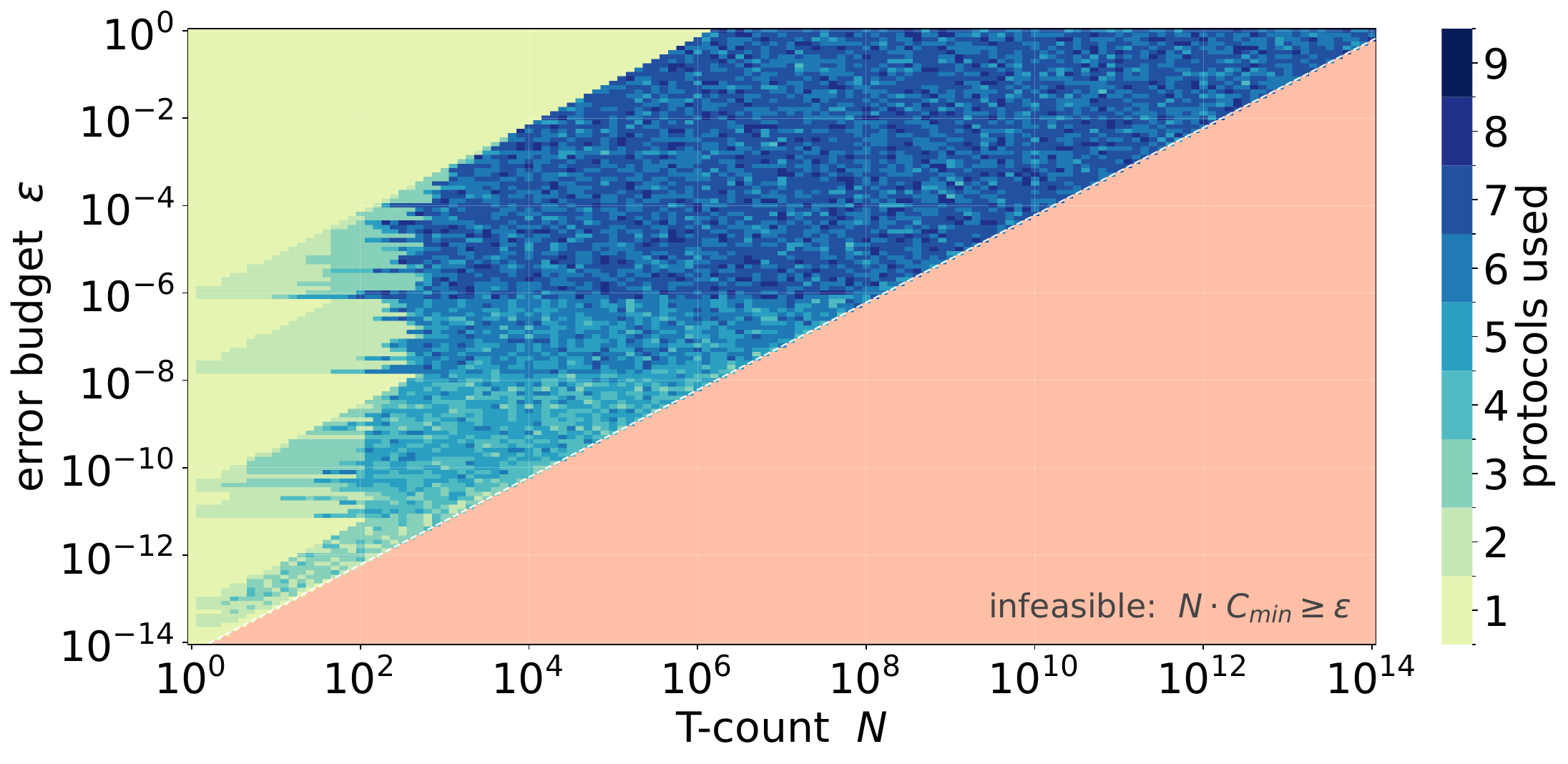}
        \vspace{-18pt}
        \caption{Factory type count
            \label{fig:fcount}}
    \end{subfigure}
    \Description{}
    \caption{Magic-state injection requirements versus target total error (\tstate{} infidelity times \tstate{} count). (a) Feasibility of cultivation alone for several benchmarks. (b) Number of distinct injection methods in the space-optimal configuration for a given fidelity target and \tstate{} count, from a constrained space-minimizing optimization over nine protocols: cultivation, \fone{}, 20-to-4, 8-to-CCZ, and cultivation-fed hybrids.}
    \label{fig:limits}
    \vspace{-12pt}
\end{figure}

\begin{figure*}[t]
    \centering
    \includegraphics[width=\linewidth]{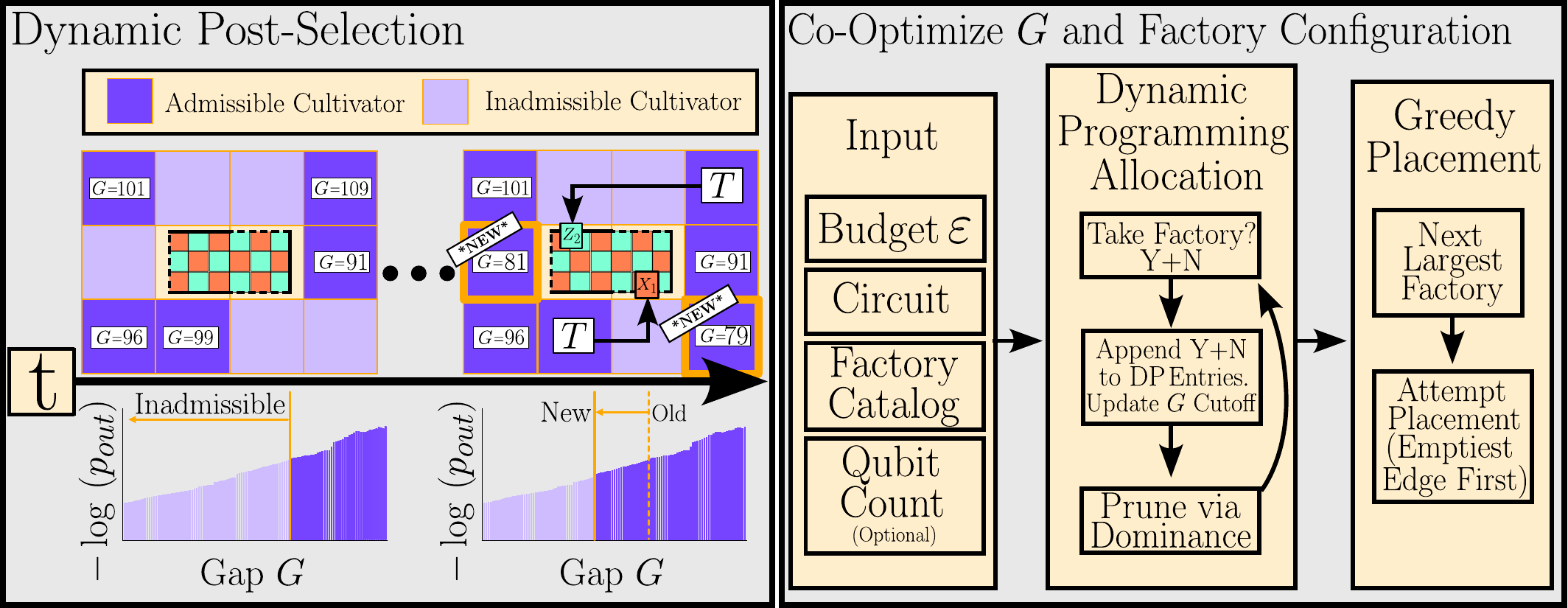}
    \Description{}
        \caption{\name{} overview. 
        \textbf{Left:} The gap cutoff is typically statically determined or baked into the cultivation process at compile time.
        Given a \terror{} budget, we propose a dynamic lattice surgery scheduler which adjusts the gap cutoff for cultivators dynamically as the budget is expended.
        If budget surplus exists, lower gap \tstate{} that could otherwise not be consumed under a static gap cutoff may be consumed to execute a PPR at the cost of additional \terror{} budget, improving execution time.
        \textbf{Right:} Given the \terror{} budget $\varepsilon$, the input circuit, a factory catalog, and optionally physical qubit count, \name{} will co-optimize gap cutoff $G$ and the allocated factory configuration in tandem.
        Allocation executes a dynamic-programming algorithm, which considers taking each entry in the factory catalog, and prunes resultant factory configuration based on dominance of \tstate{} production rate, \tstate{} quality (i.e. budget slack), and factory size, targeting one of i) minimum spacetime volume, ii) execution time, or iii) minimum space overhead.
        The dynamic-programming obtains the target objective under budget, or reports if it is not possible.
        \name{} then implements a greedy placement algorithm to realize the chosen factory configuration onto a surface code topology.
    }
    \label{fig:overview}
\end{figure*}

\section{Introduction}

Quantum computers have long promised substantial computational speedups on otherwise classically intractable problems. 
Yet, to feasibly execute relevant applications in current hardware, which suffers from noisy or otherwise imprecise operations, quantum error correction (QEC) codes must be used to encode logical information over multiple physical qubits to provide redundancy and achieve fault-tolerant quantum computation (FTQC).
One such QEC code is the surface code, which has gained widespread adoption due to its high error threshold~\cite{fowler_surface_2012}, simple and scalable logical operations~\cite{horsman_surface_2012,litinski_lattice_2018,fowler2018low}, and fast decoders~\cite{higgott,Higgott2025sparseblossom}.
However, the surface code cannot fault-tolerantly prepare the \tstate{}, a state crucial for universal QC when compiling for Clifford+$T$ gatesets.
Consequently, significant work has been dedicated towards protocols which admit \tstate{}s efficiently for use in surface codes architectures.
Magic-state preparation is therefore the most expensive operation in an FTQC in both qubits and latency, and recently there's been significant effort towards preparing them at adequate quality and pace~\cite{litinski_magic_2019,litinski_game_2019,gidney_magic_2024,toshio_star-magic_2026}.
Existing lattice surgery schedulers~\cite{hofmeyr_puremagic_2026,molavi_2025,huggins2025flasq,silva2024lsp} aim to minimize space-time volume (qubits $\times$ logical cycles) given this resource bottleneck.
\begin{figure}[htbp!]
    \centering
    \includegraphics[width=\linewidth]{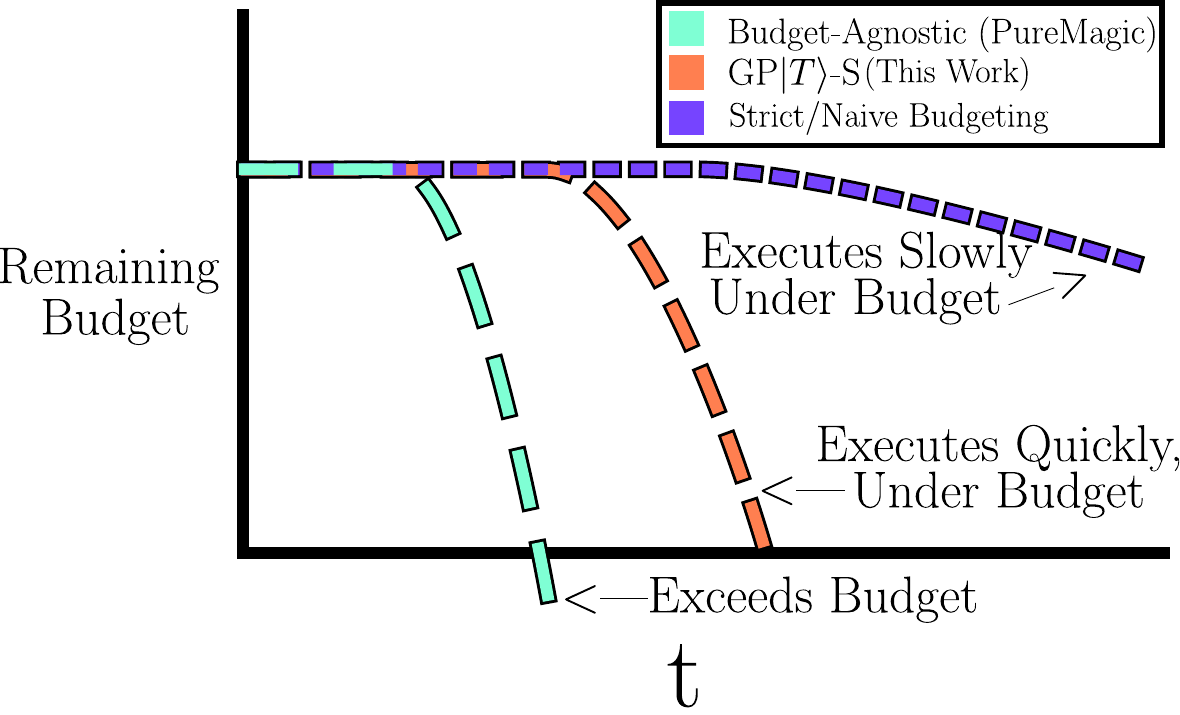}
    \Description{}
    \caption{
    Where this work sits relative to existing FTQC methods.
    Current lattice surgery schedulers are budget-agnostic, meaning that while they boast short execution times, they do not perform any resource estimation that would otherwise bottleneck practical, on-hardware execution. 
    At the other end, static allocation of a stringent gap cutoff, or of an unwieldy and large distillation factory, can both execute under budget, but at significant time cost.
    }\label{fig:execution_times}   
\end{figure}

Two methods produce \tstate{}s, and they scale very differently.
Cultivation~\cite{gidney_magic_2024} is compact and fast, growing a \tstate{} on a single logical patch.
However, its output is stochastic. 
Each preparation carries a \emph{complementary gap}, a decoder confidence score, and the fidelity of accepted states spans orders of magnitude across the gap distribution.
Cultivation reaches at best ${\approx}10^{-10}$ infidelity, only at gap cutoffs that discard most attempts.
PureMagic~\cite{hofmeyr_puremagic_2026} showed that running cultivation inside routing channels of a sparse surface code layout yields large space-time savings.
Distillation~\cite{bravyi2005universal,litinski_magic_2019} on the other hand is deterministic and reaches arbitrary fidelity by chaining factories, but a factory occupies tens of logical patches.
Moreover, the familiar ballpark \tstate{} fidelity of $p_{out}\approx 35p^3$ for a \fone{} factory holds only at distances where the factory's own logical errors are negligible.
For example, across the sizes factory allocation must consider, output infidelity varies by five orders of magnitude for \fone{} factories.
Existing schedulers fix both: a static gap cutoff chosen before execution, and hand-provisioned factories assumed to deliver one \tstate{} per cycle.
\begin{figure*}[h]
    \centering
    \includegraphics[width=\linewidth]{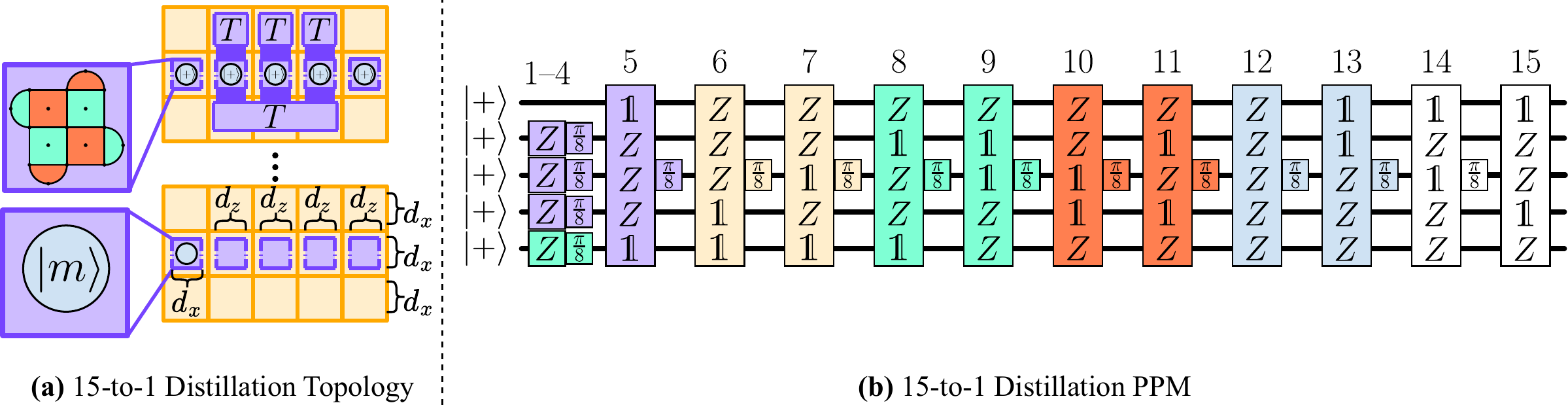}
    \Description{}
    \caption{
        \textbf{Left:} \fone{} distillation factory topology. Five logical surface code qubits occupy the middle row of the 3$\times$5 grid of tiles. The tiles in the first column have distance $d_x\times d_x$, while the remaining tiles have distance $d_z\times d_x$. Faulty $T$ rotations occur in the routing channels (top row and bottom row) according to the Pauli-product circuit seen on the right. Once this circuit completes, an output magic state $\ket m$ is produced on the left most column, ready for consumption in a Pauli-product rotation.
        \textbf{Right:} \fone{} distillation Pauli-product circuit~\cite{litinski_magic_2019}. Each product requires a single \tstate{} to execute, which are typically done via faulty-$T$ rotations, but can also be provided from external sources such as cultivation or another distillation factory. Products which share a color can be executed in parallel.
    }\label{fig:distillation}
\end{figure*}
\begin{figure}[h]
    \centering
    \includegraphics[width=\linewidth]{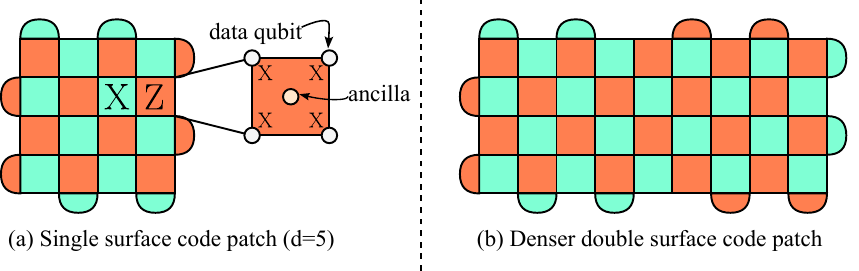}
    \Description{}
    \caption{
        (a) Left: a single distance-5 surface code patch with $X$ (red) and $Z$ (blue) stabilizers, each measured through one ancilla qubit; right: an alternative patch shape that exposes joint \emph{ports} for simultaneous logical measurements.
    }\label{fig:surface-code}   
\end{figure}

Neither method alone is generic or optimal.
If an executing program can only tolerate some finite amount of aggregate $T$-error contribution, many programs are rendered infeasible with current cultivation techniques alone (Figure~\ref{fig:cultfeas}). Distillation, at the other extreme, devotes large spatial and temporal overheads to \tstate{} filtering.
Between the extremes lies a large design space: Figure~\ref{fig:fcount} shows that the space-optimal way to reach a target fidelity mixes several methods at once.
No existing tool searches this space. 
The right cutoff depends on which factories are present, the right factories depend on the cutoff, and both depend on the program's $T$ count and error budget.

In this work, we propose Gap-Priced $\ket T$-States (\name{}), which jointly optimizes the cultivation gap cutoff, factory configuration, and lattice-surgery schedule under a total $T$-error budget. \name{} uses the complementary gap as a per-state fidelity estimate and dynamically adjusts the cutoff to convert surplus fidelity budget into speed. 
We also characterize and catalog factories by size, production rate, and budget slack, and introduce \emph{cultivating factories}.
Inspired by PureMagic~\cite{hofmeyr_puremagic_2026}, these \emph{cultivating factories} run cultivation inside factory-internal routing channels to reach fidelities otherwise requiring two-level distillation at a single-level footprint. 
\name{}'s allocator then selects Pareto-optimal configurations over qubits, production rate, and budget slack using a dynamic-programming algorithm over the factory catalog. 
When paired with dynamic cultivation post-selection, factories can provide high-quality \tstates{}, providing the necessary budget slack for low-gap cultivated \tstates{} to be consumed, which leads to significant time savings while remaining under \terror{} budget.
Across sixteen benchmarks with $T$ counts from $1.4\times10^4$ to $8.2\times10^5$ and budgets of $\varepsilon=10^{-3}$ and $10^{-5}$, \name{} outperforms cultivation-only and single-factory baselines on every benchmark without exceeding the budget, achieving a 5.5$\times$ geometric mean speedup in logical cycle count and a 3.4$\times$ geometric mean reduction in space-time volume. 
The approach also generalizes to other block-code switching schemes and factory types, supporting arbitrary factory catalogs to allocate over.

Our contributions are as follows:
\begin{itemize}
\item An estimation for cultivated-\tstate{} fidelities via the complementary gap.
\item A dynamic lattice surgery scheduler that trades fidelity for speed under a global $T$-error budget by dynamically adjusting the post-selection criteria (namely gap cutoff) for cultivators based on the remaining budget.
\item A dynamic-programming factory allocator that co-optimizes cultivation gap cutoffs and factory configurations, targeting cycle count or space-time volume minimization while remaining under \terror{} budget.
\item A distance-resolved characterization of \fone{} factories showing the $p_{out}\approx35p^3$ approximation can misestimate output fidelity by orders of magnitude.
\item A PureMagic-inspired cultivating \fone{} factory that reaches two-level-distillation fidelity at a single-level footprint via in-channel cultivation, with our catalog, simulations, and implementation.
\end{itemize}

\section{Background}
This section reviews the fault-tolerant quantum computing (FTQC) concepts that \name{} builds on: the surface code, compilation to Pauli-based computation (PBC), scheduling PBC with lattice surgery, and the two \tstate{} preparation methods, cultivation and distillation.

\subsection{The Rotated Surface Code}

The rotated surface code is a family of 2D topological quantum error correction (QEC) codes with parameters $[[n,k,d]]=[[d^2,1,d]]$: a patch of distance $d$ uses $n=d^2$ physical data qubits, encodes a single logical qubit ($k=1$), and corrects up to $(d-1)/2$ errors.
Figure~\ref{fig:surface-code} (left) shows a single patch.
All stabilizers are weight-4 (or weight-2 on the boundary) $X$ or $Z$ Pauli strings with nearest-neighbor connectivity only, which makes the code easy to implement on any grid-like architecture.
The surface code is popular in FTQC because it i) admits a high error threshold~\cite{fowler_surface_2012}, ii) has simple and scalable logical operations that can be straightforwardly implemented in hardware~\cite{horsman_surface_2012,litinski_lattice_2018,fowler2018low}, and iii) has fast decoders~\cite{higgott,Higgott2025sparseblossom}.

FTQC surface code architectures typically consist of a 2D grid of patches, and we adopt the same model in this work.
Following the extensive literature on such grids~\cite{litinski_game_2019,litinski_lattice_2018,litinski_magic_2019,beverland_surface_2022,hofmeyr_puremagic_2026,sethi_rescq_2025}, we define one logical cycle as $d$ rounds of syndrome extraction (SE).

Patches can also encode $k>1$ logical qubits (Figure~\ref{fig:surface-code}, right).
In this joint encoding, the $X$ and $Z$ observables are rearranged and joint-parity observables ($X_1X_2$, $Z_1Z_2$) are exposed at the ends/caps of the doubly-wide patch (note that this is not necessarily the case had two independent surface code patches had been joined via lattice surgery).
The total number of ports is reduced, but the representation is denser, and the exposed joint-parity observables create additional opportunities for parallelism~\cite{hofmeyr_puremagic_2026}.

\subsection{Compiling to Pauli-Based Computation}
The traditional path towards universal quantum computation requires the Clifford group plus one non-Clifford gate~\cite{nielsen2010quantum}; recent FTQC literature typically chooses the $T$ gate, $T=Z_{\frac \pi 8}=e^{i\pi/8}R_Z(\frac \pi 4)$~\cite{litinski_game_2019}.
A circuit is first transpiled to Clifford+$R_Z$, and $R_Z$ gates are then approximated via $\{H,S,T\}$ using unitary synthesis~\cite{ross_gridsynth_2016}.
The resulting circuit is then rewritten as a sequence of Pauli rotations ($R_X,R_Y,R_Z$) using the tableau method~\cite{aaronson_tableau_2004, bravyi2016pbc}.
Clifford rotations are commuted to the end of the circuit and absorbed into the final measurements~\cite{litinski_game_2019}.
Pauli byproducts arising at runtime from stochastic measurement outcomes are tracked classically in a Pauli frame via software rather than physically applied~\cite{knill_frame_2005,riesebos_frame_2017}.
What remains is a sequence of $\pi/8$ Pauli-product rotations (PPRs), each acting on several qubits in different Pauli bases, followed by Pauli-product measurements (PPMs).
This is known as Pauli-based computation (PBC).
Every PPR consumes one \tstate{}, so the \tstate{} supply is the natural bottleneck of execution under Cliford+$T$.

\subsection{Scheduling PBC with Lattice Surgery}
PPRs and PPMs are realized on the surface code grid as joint parity measurements ($XX$, $ZZ$, etc.), performed by merging and splitting patches via lattice surgery~\cite{horsman_surface_2012,fowler_surface_2012,vuillot_code_2019,litinski_lattice_2018}.
A merge combines the stabilizers of two patches, through an ancillary region between them, into a single merged surface.
Error correction runs on the merged surface, after which the ancillary region is measured out; its measurement outcomes yield the parity of the joint observable and determine any logical correction.
Which observable is measured depends on the patch edges connected by the merge.
A merge takes one logical cycle, and destructive measurements can be completed within the same cycle.
A PPR is executed by merging the target qubits with a \tstate{} through a routing ancilla and measuring the joint parity (i.e. $P\otimes Z$ between the data patch and the magic state).

Scheduling PBC therefore means assigning, in each logical cycle, the routing ancillas that connect each PPR's qubits to an available \tstate{}, subject to the grid's routing capacity.
Because surface codes cannot prepare \tstate{}s fault-tolerantly, the rate and quality at which \tstate{}s are produced and admitted dominate the schedule~\cite{litinski_magic_2019,litinski_game_2019,gidney_magic_2024,toshio_star-magic_2026}.

\subsection{$T$-State Cultivation}

\tstate{} cultivation~\cite{gidney_magic_2024} has recently gained traction because its spatial overhead is far lower than that of other preparation methods~\cite{hofmeyr_puremagic_2026,sethi_injeqt_2026,khan_architecting_2026}.
Cultivation begins by preparing a \tstate{} in a low-distance ($d\approx3,5$) color code. %
The color code is then grown by repeating check, grow, and stabilize operations, with harsh post-selection on the measurements taken along the way.
Conditioned on passing post-selection, the enlarged color code is \emph{grafted} onto a large surface code; after several rounds of SE, the state \emph{escapes} to the surface code's distance.
The resulting high-fidelity \tstate{} is then ready for consumption.

We parameterize a cultivation process by $(d_1,d_2,r_1,r_2,p)$, where $d_1$ and $d_2$ are the cultivation and escape code distances, $r_1$ and $r_2$ are the number of SE rounds at the cultivation and escape stages, and $p$ is the uniform gate error rate.
End-to-end cultivation error rates are obtained by Monte Carlo simulation using the Stim circuits provided by~\cite{gidney_magic_2024}
In this work, we default to cultivation configuration $d_1=r_1\in\{3,5\}$ and $d_2=15,\ r_2=5$.

\parah{Complementary gap}
The final post-selection step of cultivation relies on the decoder of the escape stage.
Given the observed syndrome, the decoder finds the most likely correction; it can also find the most likely correction in the \emph{complementary} logical class, i.e., the one that would flip the logical outcome.
The \emph{complementary gap} is the difference between the log-likelihood weights of these two corrections.
A large gap means the decoder is confident in its correction and the prepared state is unlikely to carry an undetected logical error; a small gap means the two hypotheses are nearly equally likely.
Cultivation therefore discards any preparation whose gap falls below a fixed \emph{gap cutoff}, and the choice of cutoff trades preparation rate for output fidelity.

\parah{PureMagic}
Because cultivation runs on a single logical patch, it can be performed directly in lattice surgery routing spaces rather than in a dedicated factory region.
PureMagic~\cite{hofmeyr_puremagic_2026} was the first scheduler to exploit this.
Under a square grid of data qubit patches with routing channels of adjustable width, PureMagic proposes performing cultivation in the routing channels rather than dedicating large swaths of qubits for factories along the peripheral of the architecture.
By treating the routing tiles of the grid as cultivation sites, the scheduler is able to launch cultivation attempts on tiles that are idle and consume successfully cultivated \tstate{}s from tile(s) adjacent to the qubits of the next PPR.
Since the latency of each attempt is stochastic, and most attempts are discarded by post-selection, PureMagic dynamically dictates which tiles should cultivate and which tiles should route PPRs.
The flexibility afforded by in-channel cultivating is capable of removing the spatial overhead of factories, cultivating or otherwise, entirely, at the cost of tying the \tstate{} supply rate to the amount of idle routing space and to the success probability of cultivation, which is fixed by a static gap cutoff.

\subsection{$T$-State Distillation}
Traditionally, the FTQC literature relies on distillation~\cite{bravyi2005universal} to prepare $T$ states fit for consumption.
Distillation applies many faulty $T$ rotations to logical surface code qubits within modular units known as factories.
A factory is defined by i) a PPM circuit consisting entirely of $Z_{\pi/8}$ rotations and ii) a surface code layout on which that circuit is executed.
The \fone{} factory for example (Figure~\ref{fig:distillation}, left) uses 15 faulty $T$ rotations, each with fault rate $p$, to output a single \emph{distilled} \tstate{} with fault rate $O(35p^3)$.
Litinski~\cite{litinski_magic_2019} shows how to implement it as a $5\times3$ grid of surface code tiles of varying, typically asymmetric code distances.
That is, its grid is not uniform; its footprint depends on the code distances $d_x,d_z$ of the individual tiles, and its production rate depends on the measurement distance $d_m$.

\section{Motivation}
\label{sec:motivation}
\subsection{$\ket{T}$-State Factory Allocation}
Current FTQC literature towards \tstate{} preparation~\cite{litinski_game_2019,gidney_magic_2024} offers a rich catalog of \tstate{} preparation methods.
Cultivation at several distances and gap cutoffs, \fone{}, 20-to-4, and 8-to-CCZ factories at many sizes, and two-level combinations of these (and even multi-level).
Moreover, recent work has found a unified numerical search method for identifying distance-2 factories~\cite{gidney_malleable_2026}.
What current FTQC literature does not offer is a tool that, given a program, an error budget, and a qubit budget, selects a configuration from that catalog.
Existing lattice surgery schedulers~\cite{silva2024lsp,molavi_2025,huggins2025flasq} take the factory count, placement, and output rate as fixed inputs; PureMagic~\cite{hofmeyr_puremagic_2026} takes the cultivation protocol and gap cutoff as fixed inputs.
The choice is left to the designer, and it is neither obvious nor separable from the schedule.

\subsection{$\ket T$-State Quality and Production Rate}
Cultivation does not produce \tstate{}s of a single fidelity.
Each successful preparation carries a complementary gap drawn from a wide distribution, and the infidelity of the prepared state spans several orders of magnitude across that distribution (see Figure~\ref{fig:gap}).
Static gap cutoff collapses this distribution, discarding states that would have been good enough for the program at hand, and it commits to a static $T$-fidelity distribution for the entire execution regardless of how much error budget remains.
Both the rate and the quality of the \tstate{} supply therefore depend on a parameter that the scheduler currently cannot adjust.

Distillation, on the other hand, is deterministic, and resource estimates commonly summarize a \fone{} factory by a single output infidelity, $p_{out}\approx35p^3$.
This describes the factory's algorithmic error and is accurate once the code distances $d_x$, $d_z$, and $d_m$ are large enough for the factory's own logical errors to be negligible.
We show this in Figure~\ref{fig:factory_oversight}.
Under uniform gate error rate of $p=10^{-3}$, the output infidelity of \fone{} factories spans nearly five orders of magnitude across sizes, and only the largest approach $35p^3$.
Moreover, the production rate scales with size as well.
$d_m$ dictates the rate at which \tstates{} are produced.
If $d_m$ isn't sufficiently large to accommodate $d_x$ or $d_z$, the output fidelity will suffer as a result.
But larger $d_m$ causes factories to output \tstate{}s at a slower rate.
Factories are therefore not a single design point but a family of them.
An allocator that sizes factories by a single formula will either over-provision or fall short of the required $T$ fidelity for successful program execution.

\subsection{Cultivation and Distillation Scaling}
Cultivation is cheap in space and fast at low fidelity, but its output saturates around $10^{-10}$ infidelity, and approaching that limit requires high gap cutoffs.
This in turn leads to longer preparation times if the cutoff is stringent enough (i.e. long-tail retry behavior~\cite{hofmeyr_puremagic_2026}).
Distillation reaches arbitrary fidelity through larger, daisy-chained factories, but each step costs a footprint many times that of a logical qubit.
Figure~\ref{fig:cultfeas} shows that many programs are infeasible with cultivation alone, while Figure~\ref{fig:fcount} shows that the space-optimal way to reach a target fidelity mixes several methods at once.
Neither method is dominant across the range of program sizes and error budgets of interest.

\subsection{Co-Optimizing Cultivation and Distillation}
The design space for \tstate{} consumption between cultivation and distillation is nontrivial.
The production rate and quality for \tstate{}s given a program are dependent on the program's $T$ count and the input error budget.
Cultivators provide access to fast and lower quality \tstate{}s, while distillation factories provide the opposite, leading to cultivators providing low quality \tstate{}s for a larger ratio of $T$ gates than distillation.
Consequently, the individual parameters for allocating cultivators and factories depend on the quality and production rate of each other, since they'll both consume a portion of the remaining error budget.
\name{} treats the gap cutoff, the factory catalog, and the schedule as one design space.
By tracking cultivated \tstate{} fidelity online through the complementary gap and budgeting it dynamically (see Figure~\ref{fig:overview} left), \name{} is able to allocate factories from a characterized catalog jointly with the gap cutoff to minimize either spacetime volume or execution time depending on the user-specified objective.

\begin{figure}[h]
    \centering
        \includegraphics[width=\linewidth]{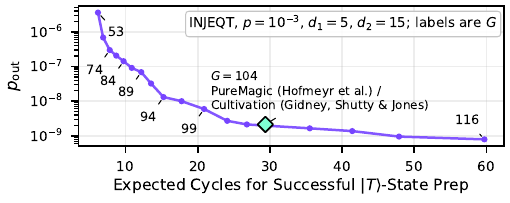}

        \includegraphics[width=\linewidth]{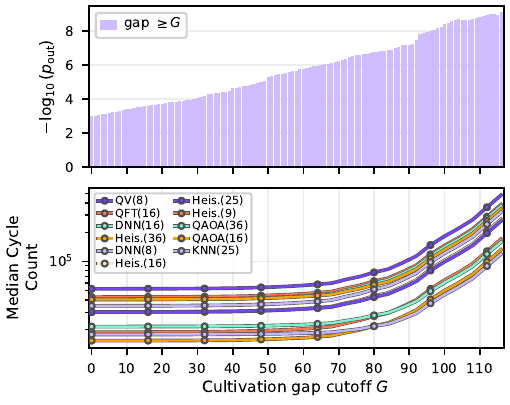}

    \caption{Complementary gap as a knob on cultivation, for $d_{\mathrm{cult}}=5$ and $d_{\mathrm{esc}}=15$ at gate error rate $p=10^{-3}$.
\textbf{Top:} Output infidelity $p_{\mathrm{out}}$ versus the expected number of logical cycles per accepted state, from the INJEQT timing model; each point is a static gap cutoff $G$ (labeled), spanning $G=53$ at 6.2 cycles to $G=116$ at 59.7 cycles, and the diamond marks the cutoff $G=104$ used by PureMagic and by Gidney et al.
\textbf{Middle:} Infidelity of the suffix aggregate the scheduler consumes, all states with gap $\geq G$, plotted as $-\log_{10} p_{\mathrm{out}}$ so higher is better; the suffix is monotone in $G$ by construction, improving from $10^{-3}$ to below $10^{-9}$ across the range.
\textbf{Bottom:} Median logical cycles versus static cutoff for 11 benchmarks (bands: interquartile range over seeds); cycle counts rise 3--9\% by $G=60$, 22--48\% by $G=80$, and 6.8--9.3$\times$ at $G=116$.
}
\label{fig:gap}
\end{figure}

\section{Tuning Cultivation with the Complementary Gap}
\label{sec:gap_fidelity}

Cultivation exposes a single parameter, the gap cutoff, that trades preparation rate for output fidelity in its post-selection criteria.
Existing schedulers fix this parameter before execution.
\name{} instead controls it during execution, raising or lowering the cutoff according to how much of the program's error budget remains.
Doing so requires two things. 
First, a way to convert a complementary gap into a fidelity estimate, and second, a policy that decides what fidelity the next \tstate{} must have.
This section provides both, then addresses a subtlety in choosing the cutoff when gap and fidelity are not monotonically related.

\subsection{From Gap to Fidelity}

The complementary gap reported by the decoder is not deterministic, depending on the decoding difficulty of the syndrome measured (i.e. how close the most likely corrections are in likelihood). 
By characterizing cultivation preparation via simulations in Stim, the gaps observed over many preparations form a distribution.
Using the publicly available end-to-end simulation data of~\cite{gidney_magic_2024}, we're able to sample such distributions directly.
The middle panel of Figure~\ref{fig:gap} summarizes the resulting gap statistics for the default \cultconfig{} configuration used throughout this work.
When modeling cultivation, we draw a gap from this distribution for each successful preparation, and the scheduler bases its runtime decisions on the drawn value.

The same simulation data also relates gap to fidelity.
Whereas the output fidelity of a distillation factory can be obtained by density matrix simulation (see~\cite{litinski_magic_2019}), cultivation is simulated with stabilizer tableau methods~\cite{aaronson_tableau_2004,gidney2021stim}, which report, for each shot, its gap and whether a logical error occurred.
Binning shots by gap yields a logical error rate (LER) per gap value.
More usefully, aggregating all shots at or above a candidate cutoff yields the average LER of the \tstate{}s that cutoff would accept.
The middle panel of Figure~\ref{fig:gap} plots this relationship for the \cultconfig{} configuration.
From it we build, for each cultivation protocol, a lookup table of \texttt{gap:fidelity} entries that lets the scheduler estimate the fidelity of a cultivated \tstate{} in constant time.

\subsection{Budgeting $T$ Error per State}

Let $\varepsilon$ be the total $T$-error budget of a program with $n_T$ $T$ gates, and let $\varepsilon_{rem}$ be the portion still unspent when the scheduler reaches the $i$th $T$ gate.
The budget allocated to the $i$th \tstate{} is the remaining budget spread evenly over the remaining gates:
\begin{equation}
\varepsilon_{T_i}=\frac{\varepsilon_{rem}}{n_T-i}.
\label{eg:init_bud}
\end{equation}
Given $\varepsilon_{T_i}$, the scheduler consults the lookup table to find the smallest gap cutoff whose accepted \tstate{}s have LER at most $\varepsilon_{T_i}$ (Figure~\ref{fig:gap}, middle).
Any cultivated state whose gap meets that cutoff is accepted immediately, with output error $\varepsilon_{out}\leq \varepsilon_{T_i}$; $\varepsilon_{rem}$ is then reduced by the gap-estimated error of the consumed state, and Eq.~\ref{eg:init_bud} is re-evaluated for the next gate.

Moving post-selection out of the cultivation protocol and into the budget-aware scheduler has two effects.
First, the scheduler sees the full fidelity distribution rather than the single accept/reject decision of a static cutoff, so it can accept states a static cutoff would have discarded and reject states a static cutoff would have passed.
Second, execution time becomes a function of the budget trajectory.
If early \tstate{}s are of higher quality than required, $\varepsilon_{rem}$ is depleted more slowly than planned, the per-state budget of the remaining gates grows, and the cutoff can be lowered, accepting a wider range of gaps and shortening the wait for each state.
Conversely, if the budget is depleted faster than planned, the cutoff rises and the scheduler waits longer for sufficiently high-fidelity states.
The scheduler thus converts surplus fidelity into speed and, when necessary, speed into fidelity, without any change to the cultivation protocol itself.

\subsection{Choosing the Cutoff when Gap Is Not Monotonic}

Eq.~\ref{eg:init_bud} assumes that fidelity improves monotonically with gap, so that the smallest cutoff meeting the budget is also the safest.
The data does not fully support this.
The \cultconfig{} results using data from~\cite{gidney_magic_2024} contain gap values at which no shot errored, as well as small dips and bumps in LER as gap increases, most likely because of finite shot counts at large gaps.
The scheduler therefore cannot simply pick the smallest gap whose per-gap LER meets the target.
Instead, it evaluates each candidate cutoff by the aggregate LER of the entire \emph{suffix} of the distribution at or above it, and the same aggregation applies recursively as the cutoff moves.

Concretely, the scheduler scans the suffixes to find the minimum cutoff whose aggregate LER satisfies the budget in expectation,
\begin{equation}
\mathbb{E} [\varepsilon_{T_i} ] = \frac {\varepsilon_{rem}} {n_T-i}.
\label{eg:expected_bud}
\end{equation}
This captures the cycle savings of a low cutoff while relying on the higher gaps later in the suffix to average out the lower ones.
For $n_T\geq10^6$ (i.e. those relevant to FTQC applications) the averaging is reliable.
However, for smaller programs, or near the end of execution, such an average may overspend, but this can be resolved by restricting no single consumption to render the remaining budget infeasible.

\section{Co-Allocating Distillation and Cultivation}
\label{sec:allocation}

The scheduler of Section~\ref{sec:gap_fidelity} tunes cultivation to the error budget, but it cannot exceed the fidelity ceiling of cultivation itself.
For programs requiring $T$ fidelity beyond that, \name{} supplements PureMagic cultivation with \fone{} distillation factories attached to the edges of the layout, and it must decide how many factories to attach, of what size, and where, jointly with the gap cutoff the scheduler should target.
This section builds up to that decision in three steps.
We first characterize \fone{} factories by size, showing that summarizing a factory by the usual $35p^3$ figure hides a wide spread in output fidelity and rate at the sizes relevant to allocation.
We then extend the factory catalog with \emph{cultivating} factories that run PureMagic cultivation inside their routing channels, reaching fidelities no single-level factory with physical $T$ injections achieves.
Finally, we formulate the allocation problem over this catalog and solve it with a dynamic programming factory allocator that returns the Pareto front over space, rate, and budget slack.

\begin{figure}[t]
    \centering
    \includegraphics[width=\linewidth]{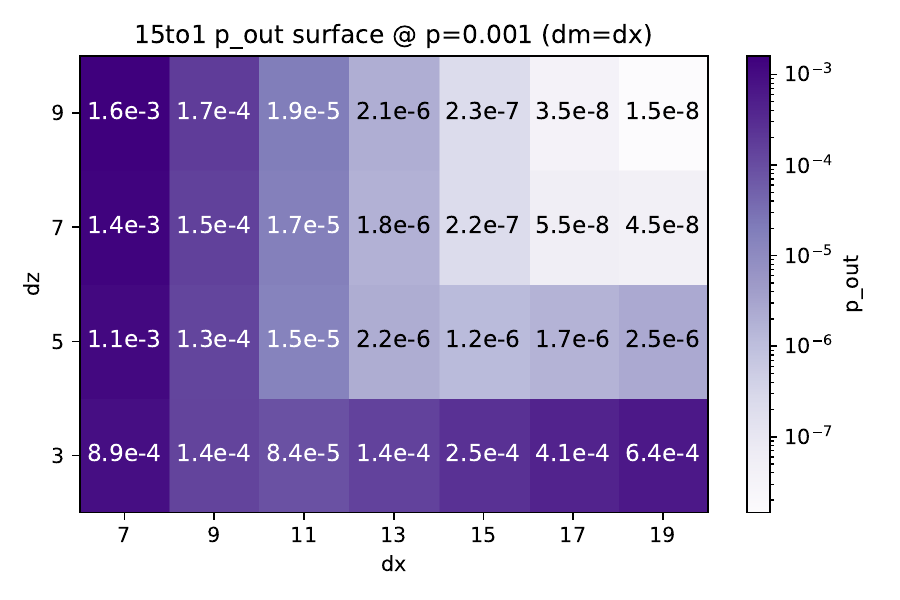}
    \Description{}
    \caption{Output infidelity $p_{\mathrm{out}}$ of \fone{} factories as a function of the tile distances $d_x$ and $d_z$, from density matrix simulation at $p=10^{-3}$ with $d_m=d_x$. Output spans nearly five orders of magnitude, from $10^{-3}$ to $1.5\times10^{-8}$. The $35p^3=3.5\times10^{-8}$ approximation is approached only for $d_x\geq17$ and $d_z\geq7$; at $d_z=3$, increasing $d_x$ does not improve the output at all.}
    \label{fig:factory_oversight}
\end{figure}

\subsection{Characterizing \fone{} Factories}

Resource estimates commonly model a \fone{} factory as a black box turning 15 faulty $T$ rotations of fault rate $p$ into one \tstate{} of infidelity $O(35p^3)$.
This is accurate when the factory's code distances are large enough that its own logical errors are negligible.
Factory allocators, however, must compare factories across sizes, including small ones where that regime does not hold.
Figure~\ref{fig:factory_oversight} shows density matrix simulations of \fone{} factories across a range of sizes, all at a uniform gate error rate of $p=10^{-3}$ with the same rate for the faulty $T$ rotations.
Output infidelity spans nearly five orders of magnitude.
The largest factories ($d_x\geq17$, $d_z\geq7$) land at $1.5$--$4.5\times10^{-8}$, agreeing with $35p^3$, while smaller ones are dominated by their own logical errors.
Factories must therefore be sized to the fidelity the program needs, and below saturation, output fidelity decouples from input fidelity.

Production rate depends on size as well.
Because \fone{} factories are more sensitive to $Z$-type errors, we set the measurement distance $d_m=d_x$ for every factory in this work, so the number of cycles between outputs grows with $d_x$.
Fidelity and rate are therefore both functions of factory size, and a factory catalog must record both.

\subsection{Cultivating Factories}

The Heisenberg-36 Hamiltonian evolution benchmark requires $O(10^5)$ $T$ gates (assuming gridsynth~\cite{ross_gridsynth_2016} tolerance of $10^{-8}$), so even at a generous budget $\varepsilon=10^{-3}$ each \tstate{} may contribute at most ${\sim}10^{-9}$.
Standalone \fone{} factories saturate near $10^{-8}$ (Figure~\ref{fig:factory_oversight}), an order of magnitude short.
Moreover, \cultconfig{} cultivation reaches $10^{-9}$ only at high cutoffs and hence slowly.
The traditional remedy is a two-level scheme chaining \fone{} or 20-to-4 factories~\cite{litinski_game_2019,litinski_magic_2019}, at a large spatial cost.

We instead run PureMagic cultivation inside the routing channels of a \fone{} factory, replacing its 15 faulty physical $T$ injections with cultivated \tstate{}s.
Recall the \fone{} factory layout has ten routing ancilla tiles, five in the top row and five in the bottom (see Figure~\ref{fig:distillation}).
Execution for a cultivating \fone{} variant proceeds as follows.
First, five \tstate{}s are cultivated in any of the ten tiles and injected to perform rotations 1-5 simultaneously in $d_m$ cycles each.
For each of the remaining ten rotations, cultivation attempts run in parallel in the top and bottom channels; as soon as one succeeds, the corresponding injection is performed in $d_m$ cycles.
The expected (or worst-case) execution time is therefore $6d_m+6\Delta$, where $\Delta$ is the expected (or worst-case) time for a round of parallel cultivation attempts.
Because cultivation occupies the routing channels that the factory already contains, no additional ancilla space is needed, provided the escape distance does not exceed $\min(d_x,d_z)$; the larger footprints reported in Table~\ref{tab:factories} for cultivating factories are due solely to their larger surface code distances. %

\begin{figure}[t]
    \centering
    \includegraphics[width=\linewidth]
        {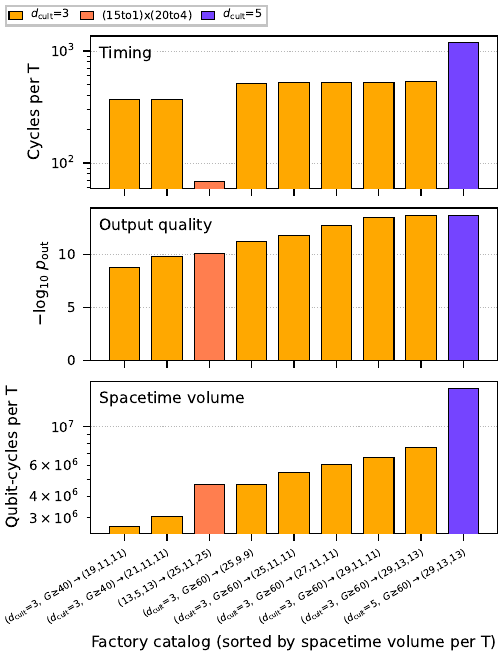}
    \caption{Per-\tstate{} metrics for the factory catalog at
        $p=10^{-3}$, sorted by space-time volume per output state:
        cycles per \tstate{} (top), output infidelity (middle), and
        qubit-cycles per \tstate{} (bottom). Orange bars are cultivating
        \fone{} factories with $d_{\mathrm{cult}}=3$, labeled by
        $(d_x,d_z)$ and gap cutoff $G$; red is a two-level
        (15-to-1)$\times$(20-to-4) scheme; purple is a bare
        $d_{\mathrm{cult}}=5$ cultivator. Cultivating factories are
        slower per state than the two-level scheme but reach several
        orders of magnitude lower infidelity at comparable space-time
        volume, and dominate the bare cultivator on all three metrics.}
    \Description{}
    \label{fig:culti_vs_distill}
    \vspace{-8pt}
\end{figure}

\begin{figure}[t]
    \centering
    \includegraphics[width=\linewidth]
        {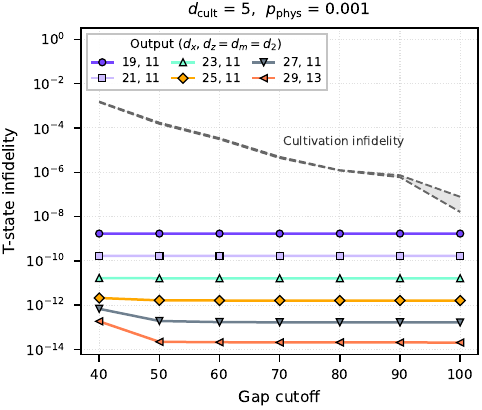}
    \caption{Infidelity of cultivate-then-distill \tstate{}s versus gap
        cutoff, compared with bare cultivated \tstate{}s at escape
        distances $d_2\in\{11,13\}$. Although the input states improve
        with the cutoff, the factory output does not, so a higher cutoff
        should be spent on distillation time rather than fidelity.}
    \Description{}
    \label{fig:l1_cultivation_tradeoff}
    \vspace{-8pt}
\end{figure}

Table~\ref{tab:factories} shows that cultivated \tstate{}s at escape distance 15 with high gap cutoffs already approach the fidelity of a \tstate{} distilled at $d_x=d_z=19$, $d_m=9$.
Feeding such states into that factory yields output infidelity between $10^{-8}$ and $10^{-9}$, at the cost of longer execution.
The output saturates near $10^{-9}$ regardless of how far the gap cutoff is raised, meaning at $p=10^{-3}$, idling error inside the factory dominates once the input states are good enough (see Figure~\ref{fig:l1_cultivation_tradeoff}).
Raising the factory distances to the level-2 values of lifts this ceiling.
Within a fixed factory size, however, a higher gap cutoff on the inputs buys little fidelity and should instead be spent on speed.
Cultivating factories quickly reach fidelities that otherwise require two-level distillation, at a single-level footprint and comparable space-time volume per state, though at a lower output rate than a two-level scheme (see Figure~\ref{fig:culti_vs_distill}), which is why they are essential members of the catalog (Table~\ref{tab:factories}).

\begin{table}[t]
  \centering
  \footnotesize
  \caption{Cultivating factories passing the pre-budget filter at $p=10^{-3}$.  Distilled
  rows are shown at their saturation-onset gap cutoff; within a $d_x$ class only
  nondominated factories are listed for brevity. 
  That is, a factory is dropped only if another in
  the same $d_x$ class is at least as good in output infidelity, output rate, and qubit count simultaneously.  Incomparable factories are retained.  $d_z = d_m$ throughout, and
  $d_{\mathrm{cult}}$ is the first-stage cultivation distance.  $\tau$ is in SE
  cycles; Qubits is the $2d_x\cdot d_z-1$ mapped area.}
  \label{tab:factories}
  \begin{tabular}{@{}rrrr r r r@{}}
    \toprule
    $d_x$ & $d_z$ & $d_{\mathrm{cult}}$ & $G$ & $p_{\mathrm{out}}$ & $\tau$ & Qubits \\
    \midrule
    \multicolumn{7}{@{}l}{\textit{Cultivated 15-to-1}} \\
    19 & 11 & 3 & 40 & $1.70{\times}10^{-9}$ & 372.6 & 7167 \\
    21 & 11 & 3 & 40 & $1.69{\times}10^{-10}$ & 372.6 & 8175 \\
    23 & 11 & 3 & 60 & $1.63{\times}10^{-11}$ & 525.2 & 9231 \\
    25 & 9 & 3 & 60 & $5.93{\times}10^{-12}$ & 513.5 & 9135 \\
    25 & 11 & 3 & 60 & $1.62{\times}10^{-12}$ & 525.2 & 10335 \\
    27 & 11 & 3 & 60 & $1.77{\times}10^{-13}$ & 525.2 & 11487 \\
    29 & 11 & 3 & 60 & $3.56{\times}10^{-14}$ & 525.2 & 12687 \\
    29 & 13 & 3 & 60 & $2.14{\times}10^{-14}$ & 537.2 & 14079 \\
    \midrule
    \multicolumn{7}{@{}l}{\textit{Bare cultivators}} \\
    15 & 15 & 5 & 100 & $2.29{\times}10^{-8}$ & 1522.6 & 449 \\
    15 & 15 & 5 & 110 & $6.78{\times}10^{-9}$ & 2739.2 & 449 \\
    \bottomrule
  \end{tabular}
\end{table}

\subsection{The Allocation Problem}

Given the catalog, an allocation fixes four things: the size of each factory, its timing, its placement along the layout edges, and the gap cutoff the budgeting scheduler should target.
Size determines output fidelity (Figure~\ref{fig:factory_oversight}) and bounding box.
Normally, allocators would also have to consider choice of $d_m$, however, we fix $d_m=d_x$, so production rate is also tied to factory size.
Placement determines the average length of the routing ancilla needed to bring a factory's \tstate{} to the qubits of a PPR, and hence the routing pressure the factory exerts.
Longer ancillas in particular monopolize channels that other operations may need.
Higher-fidelity factories are what make an otherwise infeasible program feasible, and exhibit the highest potential for program speedups by buying budget slack for the cultivators to use.
The gap cutoff determines both the rate and the fidelity of cultivated \tstate{}s, and it's coupled to the factory choice due to this provisional budget slack afforded by larger factories.
The rate and fidelity of cultivation must therefore be optimized together with the factory set rather than after it.

Each \fone{} factory in the catalog is nominally characterized by $d_x,d_z,d_m$, but for allocation these reduce to its output infidelity $p$ and its mean output interval $\tau$ in cycles.
Placement adds four attributes: $l$, the length of the interval the factory occupies along an edge; $q$, its physical qubit cost ($6d_x(d_x+4d_z)-15$ for a \fone{} body); $\gamma$, the tile on the edge through which it delivers its output; and $\kappa=\pm1$, the side of $\gamma$ on which the factory body lies.
A factory is thus $i=(p,\tau,l,q,\gamma,\kappa)$, occupying the interval $[\alpha,\beta]$ with $\alpha=\min(\gamma,\gamma+\kappa l)$ and $\beta=\max(\gamma,\gamma+\kappa l)$.
We assume the shorter edge of the factory, the one containing the output qubit, is the edge attached to the layout, and we place no constraint on how far the factory extends outward.
We denote the set of all candidate factories, over all sizes and positions, by \setp{}.

Two candidates $i,j\in\mathcal{P}$ conflict if their intervals overlap.
A configuration is valid if it contains no conflicting pair and at most one factory per output port; the set of valid configurations is the ground set $X$.
A factory may overhang either end of an edge. 
We record this with flags $c_L,c_R$ and resolve the resulting corner conflicts when edges are combined.

\subsection{Solving for the Pareto Front}

Our catalog offers a rich spread of fidelities, rates, and sizes, and the right choice depends on whether the designer wants to minimize time, space, or space-time volume under a fixed qubit budget.
Rather than commit to one objective just yet, we compute the \emph{exact} Pareto front of valid configurations with a dynamic programming algorithm.

Let $f$ be a 2D array whose rows index candidate factories and whose columns index the number of factories in a configuration.
$f_{i,j}$ holds the set of non-dominated valid configurations that include factory $i$ and contain $j$ factories.
Storing every configuration would be intractable, so each $f_{i,j}$ retains only configurations that are non-dominated with respect to three statistics: combined physical qubit count $q$ (lower is better), combined factory and cultivation production rate $r$ (higher is better), and infidelity-weighted output rate $\varphi$ (lower is better).
One configuration dominates another if it is at least as good in all three and strictly better in one; dominated entries are deleted on insertion, and tied or incomparable entries are kept (Algorithm~\ref{alg:plugins-op}, Appendix~\ref{app:exact-dp}).
Each entry also carries its overhang flags $c_L,c_R$.

Two operations build $f$ for a single edge.
\extend{} adds factory $i$ to every configuration in a row, shifting it one column right and appending $i$.
Because it draws only from $X$, the result is valid by construction.
\merge{} combines the previous row with the newly extended one by taking the column-wise union and applying the dominance update element-wise.
Both procedures, and the edge scan that applies them to produce an edge's $f$, are given in full in Appendix~\ref{app:exact-dp} (Algorithms~\ref{alg:plugins-op} and~\ref{alg:edgescan-op}).
Because the PureMagic layouts considered here are rectangular, either one edge's $f$ is replicated on all four sides or two edges' $f$ are mirrored.

\begin{figure}[ht]
    \centering
    \includegraphics[width=\linewidth]{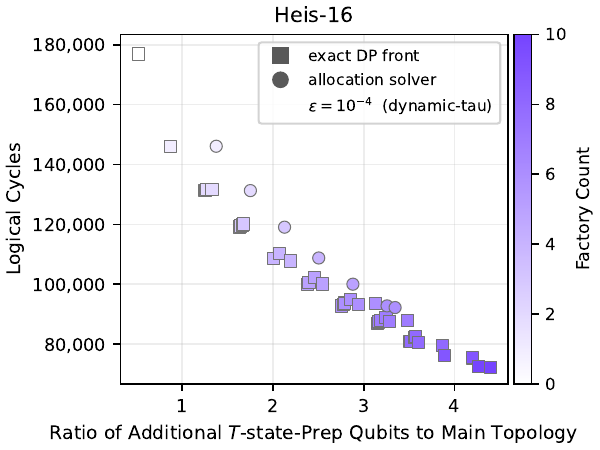}
    \Description{}
\caption{Measured Pareto fronts for Heis-16 at $\varepsilon = 10^{-4}$, colored by seated factory count: squares are the exact-DP front (37 points, 0--10 factories), circles the relaxed allocator (7 points, 1--6).
At matched factory count the allocator matches the exact front in time ($0.5\%$ \emph{fewer} cycles on average) but over-provisions space by ${\approx}$12.8k qubits, $0.45\times$ the compute core.
}
    \label{fig:pareto_shift}
\end{figure}

Algorithm~\ref{alg:ringjoin-op} (Appendix~\ref{app:exact-dp}) joins the per-edge tables around the ring of four edges.
Adjacent edges are unioned to obtain the same three statistics as in the per-edge construction, configurations whose corner flags conflict are discarded, and the dominance relation is applied to the survivors.
Feasibility against $\varepsilon$ is checked on a precomputed ladder of gap cutoffs from the lookup of Section~\ref{sec:gap_fidelity}.
A cutoff $G$ is feasible if the rate-weighted infidelity of \tstates{} from both cultivation and distillation sources is as good as $\varepsilon/n_T$.
The lowest feasible cutoff $G$ meeting this criteria is taken, converting leftover budget into supply rate.
The result is the Pareto front over $(q,r,\varphi)$, from which the fastest, the smallest, or the most space-time-efficient configuration can be read off for any qubit budget.
Figure~\ref{fig:pareto_shift} shows the measured front this program yields for Heis-16 at \terror{} budget $\varepsilon=10^{-4}$, alongside the front produced by the relaxed allocator of Section~\ref{sec:method}.
The relaxation tracks the exact front closely in time while overprovisioning space, at a solve cost orders of magnitude lower.

\begin{figure}[ht]
    \centering
    \includegraphics[width=\linewidth]{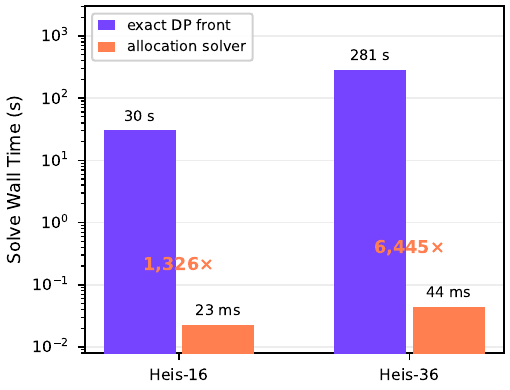}
    \Description{}
\caption{
Solver wall time at $\varepsilon=10^{-4}$, the budget of Figure~\ref{fig:pareto_shift}.
On Heis-16, exact dynamic programming (exact-DP) took 30.4 s and the allocator 23 ms, a 1,326$\times$ speedup.
On Heis-36, exact DP took 281 s and the allocator 44 ms, a 6,445$\times$ speedup.
The allocator reaches configurations that perform similarly to the exact solution at a solve cost three to four orders of magnitude smaller.
}
    \label{fig:solver_timings}
\end{figure}

\section{Methodology}
\label{sec:method}

\subsection{Implementation}

\name{} is implemented in Rust as an extension of the PureMagic lattice surgery scheduler~\cite{hofmeyr_puremagic_2026}.
It combines three components: the budget-driven gap cutoff of Section~\ref{sec:gap_fidelity}, the factory catalog and allocator of Section~\ref{sec:allocation}, and the PureMagic scheduler, which we extend to draw \tstate{}s from factories as well as from cultivation sites.
Cultivation is modeled statistically rather than simulated at runtime.
For each successful preparation we draw a complementary gap from the \cultconfig{} distributions of~\cite{gidney_magic_2024}, as processed by INJEQT~\cite{sethi_injeqt_2026}, and look up the corresponding fidelity.
Factory output fidelities and rates come from the density matrix simulations of Section~\ref{sec:allocation}.

We use a hybrid of the PureMagic layout and a bus layout, which consists of a square grid of double-data patches separated by single routing channels, with factories attached along the outer edges and their outputs delivered onto the routing ring.
Single channels and double-data patches are chosen for compactness, so that the benchmark cores fit in the qubit counts of Table~\ref{tab:benchmarks}.
Each PPR is routed as a Steiner tree over the routing channels connecting the participating data patches to the tile holding the consumed \tstate{}, using the classical approximation of~\cite{kou1981fast}.
Programs are compiled to PBC at Pauli width 1, following the parallelism argument of PureMagic, and rotations are synthesized with Gridsynth~\cite{ross_gridsynth_2016} at tolerance $10^{-8}$ (the PureMagic default).
\begin{algorithm}[t]
\caption{
Allocation solver. A bounded knapsack over factory-type
multi-sets, bucketed by instance count up to the port cap $P$.
Each vector and instance carries qubit cost $q$, output rate $r$, cumulative tangent length $t$,
and the infidelity-weighted output rate
$\varphi = \sum_f r(f)p_{\mathrm{out}}(f)$;
minimizing $\varphi$ leaves the most error budget for cultivated states.
\textsc{ParetoPrune} prunes each bucket to its
set of mutually non-dominated vectors under $(q\min, r\max, \varphi\min)$.
The perimeter guard $v.t + t(f) > P\cdot d$ is
monotone under extension, so it prunes entire multiplicity subtrees.
The returned $v^\ast$ is a starting point for the port placement to work
from. Feasibility of an operating point reuses the exact engine's suffix machinery, which converts $\varphi$ into the remaining budget by fixing the gap cutoff low enough while maintaining feasibility ($\pi$ selects among feasible points).
}
\label{alg:allocator}
\begin{algorithmic}[1]
\Require catalog items $\mathcal{F}$; port count $P$; main distance $d$;
budget $\varepsilon$; $T$ count $n_T$; policy $\pi$
\Ensure selected allocation $v^\ast$ and its operating point
$\mathrm{op}^\ast$
\State $\mathcal{F} \gets \{f \in \mathcal{F} :
  p_{\mathrm{out}}(f) < \varepsilon / n_T\}$
  \Comment{per-$T$ eligibility}
\State $B[0] \gets \{\mathbf{0}\}$;\ $B[k] \gets \emptyset$ for
  $k = 1 \dots P$ \Comment{buckets by instance count}
\For{$f \in \mathcal{F}$ ordered by $(q \uparrow, r \downarrow,
    \varphi \uparrow)$} \Comment{dominators first}
  \For{$k = 1 \dots P$}
    \For{$v \in B[k-1]$}
      \If{$v.t + t(f) > P \cdot d$} \textbf{continue}
        \Comment{perimeter prune}
      \EndIf
      \State add $v + (q(f), r(f), \varphi(f), t(f))$ to $B[k]$
    \EndFor
    \State $B[k] \gets \Call{ParetoPrune}{B[k]}$
  \EndFor
\EndFor
\State $V \gets \Call{ParetoPrune}{\bigcup_k B[k]}$
\For{$v \in V$}
  \State $G^\ast(v) \gets \min\{G : (r_C(G)\bar{p}_C(G) + v.\varphi)
    / (r_C(G) + v.R) \le \varepsilon / n_T\}$
    \Comment{lowest feasible cutoff}
\EndFor
\State $\mathcal{A} \gets \{(v, G^\ast(v)) : G^\ast(v)$ exists$\}$
\State \Return $\pi(\mathcal{A})$
\end{algorithmic}
\end{algorithm}

\subsection{Allocation Algorithms}

\name{} provides two allocators that trade solve time for optimality.
They are independent solvers over the same catalog and budget machinery, and we compare their produced fronts in Figure~\ref{fig:pareto_shift} and their associated runtimes in Figure~\ref{fig:solver_timings}.

The first is the dynamic program of Section~\ref{sec:allocation}.
Exact-DP scans the concrete candidate placements of every edge, joins the edges around the ring, and returns the exact Pareto front over qubit count $q$, production rate $r$, and infidelity-weighted output rate $\varphi$, together with a witness placement set for every point (Appendix~\ref{app:exact-dp}).
Because it enumerates placements and their conflicts, its cost grows with layout perimeter and catalog size, from tens of seconds on the smaller cores to minutes on the larger ones per budget.
Where the budget is infeasible, it proves so by returning an empty front at comparable cost.
The second allocator (Algorithm~\ref{alg:allocator}) drops the placement dimension from the search and restores it afterward.
Its solver is a bounded knapsack over factory-type multisets against the total edge length, bucketed by instance count (i.e., the number of additional \tstate{} sources) up to the port cap, with per-bucket dominance pruning over qubit cost $q$, output rate $r$, and the infidelity-weighted output rate $\varphi$.
The only geometry retained is i) minimal tangent lengths of the factories, mapped onto the knapsack, and ii) the total perimeter length, which prunes whole multiplicity subtrees.
Dropping placement is justified empirically.
Sweeping factory positions with the factory set held fixed changes cycle counts only marginally on programs from 8 to 433 logical qubits (Figure~\ref{fig:placement_bars}, Appendix~\ref{app:placement}).
Feasibility of each surviving multiset is evaluated with the same budget machinery as the exact engine, and a selection policy (minimum qubits, minimum forecast cycles, or minimum of their product) picks one point.
A greedy realizer then assigns concrete ports. 
Factories are seated in decreasing order of blocked tangent length, each at the first conflict-free candidate placement on the edge with the most free ports.
When no port admits a factory, up to three displacements evict an already-seated factory, relocate it to another edge, and drop it only if the newcomer then seats, undoing the eviction otherwise.
The realized layouts are what the scheduler executes.
Because the relaxation ignores packing until realization, its selections over-provision space (by a near-constant half of a compute core on the fronts of Figure~\ref{fig:pareto_shift}) while matching the exact front's cycle counts at equal factory count, at a solve cost roughly 3-4 orders of magnitude lower (see Figure~\ref{fig:solver_timings}).

From the Pareto front, we extract three configurations for each benchmark and budget.
The \emph{minimum-feasible} configuration is the smallest factory and gap-cutoff configuration (FGC) that can meet the budget at all.
The \emph{minimum-space-time} configuration minimizes the product of qubit count and logical cycle count.
The \emph{minimum-cycle} configuration is the FGC beyond which additional factories no longer reduce logical cycle count, i.e., the point at which routing congestion rather than \tstate{} supply bounds execution.
We also record, for each code distance and gate error rate, whether a given budget is feasible at all.
\begin{table*}[ht!]
  \centering
  \begin{tabular*}{\linewidth}{@{} c c c c c c}
    \toprule
    Benchmark & Cliffords & $n_T$ & Depth & Num. qubits & Application \\
    \midrule
    ising-\{34, 420\}      & \{758, 28334\}  & \{13544, 168504\} & \{984, 1307\}    & \{58370, 598517\}  & Hamiltonian evolution \\
    qaoa-\{16, 100\}       & \{659, 3878\}   & \{22292, 152520\} & \{7744, 25447\}  & \{28287, 150864\}  & Combinatorial optimization \\
    knn-\{25, 129\}        & \{870, 5665\}   & \{35883, 171983\} & \{13803, 56228\} & \{49390, 196213\}  & Distance-based inference \\
    heisenberg-\{16, 225\} & \{766, 34076\}  & \{45854, 815386\} & \{11157, 54330\} & \{28287, 347975\}  & Hamiltonian evolution \\
    qv-8                   & 1143            & 48614             & 7505             & 16164              & Randomized benchmarking \\
    qft-\{16, 100\}        & \{2650, 41429\} & \{65107, 667498\} & \{12140, 85881\} & \{28287, 150864\}  & Phase-estimation subroutine \\
    wstate-\{76, 118\}     & \{1395, 2140\}  & \{70690, 108748\} & \{24441, 37416\} & \{122577, 179600\} & Multipartite entanglement \\
    adder-\{64, 118\}      & \{3995, 7792\}  & \{76415, 147107\} & \{23044, 45766\} & \{99229, 179600\}  & Arithmetic primitive \\
    qugan-71               & 4981            & 157252            & 36031            & 110903             & Adversarial training circuit \\
    \bottomrule
  \end{tabular*}
  \caption{Benchmark suite, with size variants union-set and labeled by their
  qubit counts. Braced entries are $\{$smaller, larger$\}$ in the order given by
  the benchmark name. Cliffords and Depth are the Clifford-product count and PPM
  layer count reported by the circuit loader; $n_T$ is $T$ rotations; Core qubits
  is the compute-core physical footprint at main code distance $d = 15$,
  excluding factories. Ordered by $n_T$ of the smaller instance.}
  \label{tab:benchmarks}
\end{table*}

\subsection{Baselines}

We compare against two baselines.
The \emph{cultivation-only} baseline is PureMagic with a static gap cutoff of $G=104$, escape distance $d_{\mathrm{esc}}=15$, and cultivation distance $d_{\mathrm{cult}}=5$.
This configuration was chosen because it is the least restrictive cutoff at which cultivation alone attains the required per-\tstate{} fidelity for the smaller benchmarks.
It corresponds to the diamond in the top panel of Figure~\ref{fig:gap}.
For every benchmark that cultivation alone cannot satisfy, the \emph{single-factory} baseline adds one instance of the smallest catalog factory whose output fidelity meets the per-\tstate{} budget, with no cultivation, representing the static provisioning that existing schedulers assume.
Both baselines run on the same layout, routing, and compiler settings as \name{}.

\subsection{Benchmarks}

Table~\ref{tab:benchmarks} lists the benchmark suite, selected from the BenchPress~\cite{nation2025benchmarking} suite.
It spans Trotterized Hamiltonian evolution (Ising and Heisenberg models), combinatorial optimization (QAOA), distance-based inference (KNN), the quantum Fourier transform, quantum volume circuits, W-state preparation, arithmetic (adders), and an adversarial training circuit (QuGAN), each in a smaller and a larger instance distinguished by qubit count.
$T$ counts range from $1.4\times10^4$ to $8.2\times10^5$, so that at a fixed total error budget the per-\tstate{} fidelity requirement spans nearly two orders of magnitude across the suite, covering both the regime cultivation alone can serve and the regime that requires factories.
Core qubit counts are reported at main code distance $d=15$ and exclude factories.

We evaluate every benchmark at two total $T$-error budgets, $\varepsilon=10^{-3}$ and $\varepsilon=10^{-5}$, chosen to exercise both ends of budget pressure.
At $\varepsilon=10^{-3}$ the per-\tstate{} requirement $\varepsilon/n_T$ lies between $10^{-9}$ and $10^{-7}$, within reach of cultivation alone for all but the largest instances, so the question is how much speed the surplus budget can buy.
At $\varepsilon=10^{-5}$ the requirement lies between $10^{-11}$ and $10^{-9}$, beyond cultivation for every instance, so factories are mandatory and the question is how few of them, and how much cultivation alongside them, suffice.
\begin{figure*}[htbp!]
    \centering
    \begin{subfigure}[t]{\linewidth}
    \centering
        \includegraphics[width=0.875\linewidth]{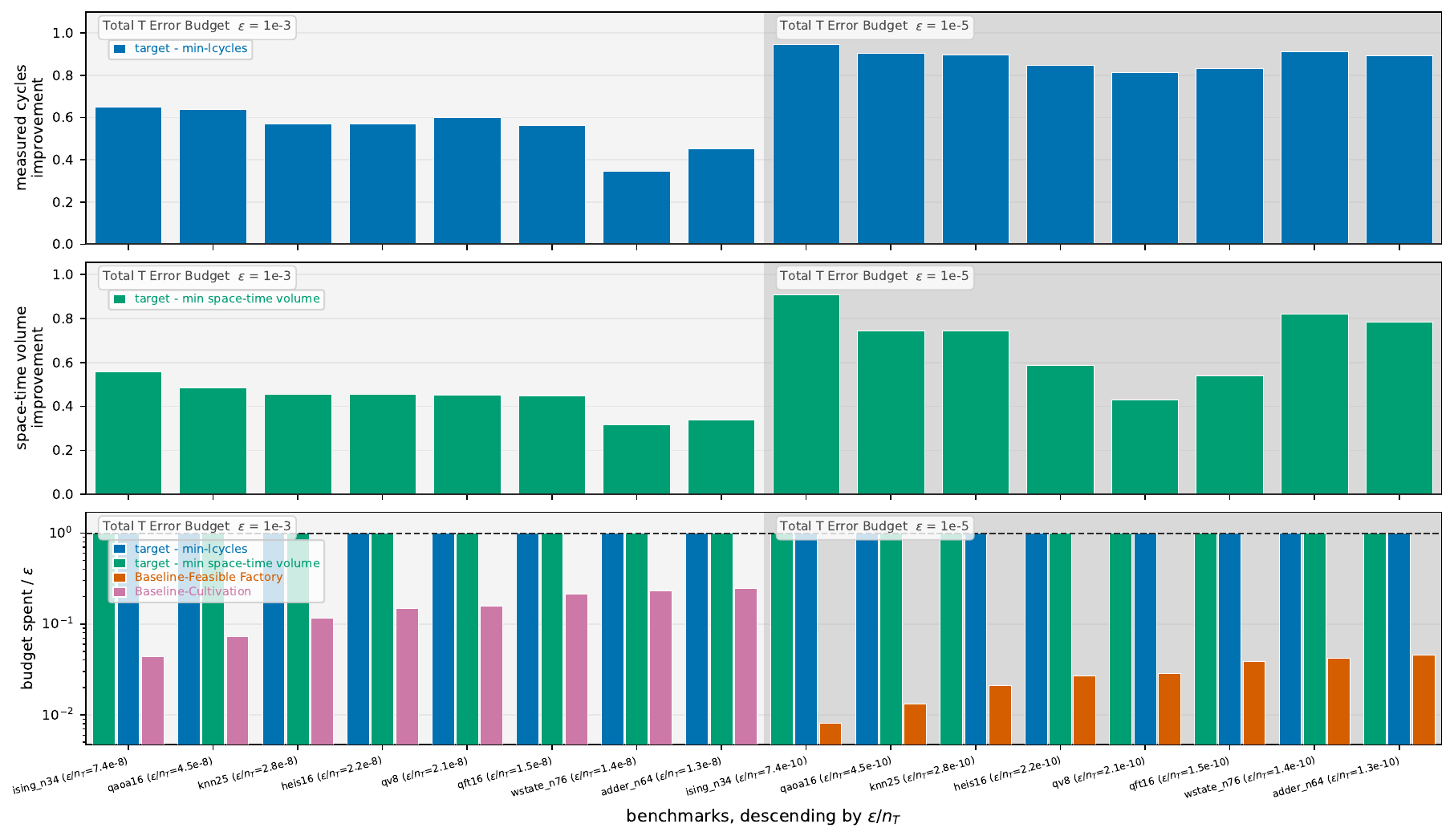}
        \caption{Small benchmarks}
        \label{fig:results_small}
    \end{subfigure}\hfill
    \vspace{8pt}
    \begin{subfigure}[t]{\linewidth}
    \centering
        \includegraphics[width=0.875\linewidth]{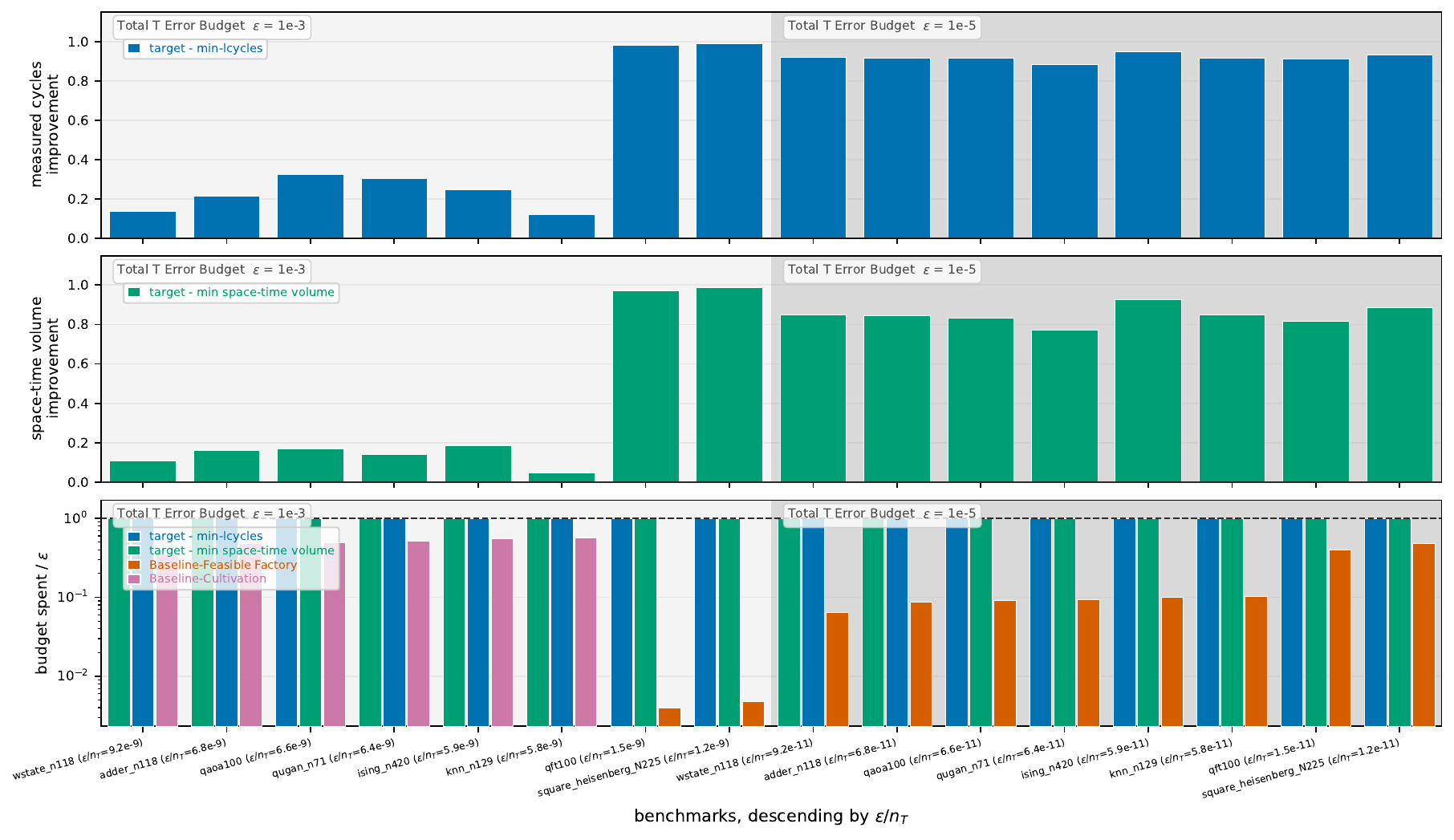}
        \caption{Large benchmarks}
        \label{fig:results_large}
    \end{subfigure}
    \Description{}
    \caption{\name{} against the baselines of Section~\ref{sec:method} on the (a) small and (b) large benchmark instances, at total $T$-error budgets $\varepsilon=10^{-3}$ (left half of each panel) and $\varepsilon=10^{-5}$ (right half). Benchmarks are ordered by decreasing per-\tstate{} budget $\varepsilon/n_T$. Top row: relative reduction in logical cycle count for the minimum-cycle configuration. Middle row: relative reduction in space-time volume for the minimum-space-time configuration. Bottom row: fraction of $\varepsilon$ actually spent by each configuration (log scale); the dashed line is the budget. The baseline is cultivation-only where it is feasible (pink) and the single smallest feasible factory otherwise (orange).}
    \label{fig:spacetime_volume_tradeoffs}
\end{figure*}

\section{Experimental Results}
\label{sec:results}

Figure~\ref{fig:spacetime_volume_tradeoffs} compares \name{} against the baselines across the sixteen benchmarks of Table~\ref{tab:benchmarks} and both budgets, 32 benchmark--budget pairs in all.
For each pair we report the minimum-cycle and minimum-space-time configurations from the Pareto front, each against the baseline on its own objective; improvement is $1-\text{\name{}}/\mathrm{baseline}$, so 0.5 halves the metric.
The bottom rows show the fraction of $\varepsilon$ each configuration consumed.
Aggregate speedups quoted in this work are $\mathrm{baseline}$/\name{}, combined across pairs by geometric mean.

\subsection{Cycle Count and Space-Time Volume}

Our work improves on the baseline in every one of the 32 pairs, on both metrics.
Logical cycle count is reduced by 69\% on average (minimum 12\%, maximum 99\%), and space-time volume by 58\% on average (minimum 5\%, maximum 99\%).
At $\varepsilon=10^{-5}$, where no benchmark is feasible with cultivation alone, gains are large and uniform, yielding improvements between 81--95\% in cycles and 43--93\% in volume for small benchmarks, and between 88--95\% and 77--93\% for cycles and volume for large benchmarks, respectively.
At $\varepsilon=10^{-3}$, where cultivation is feasible for all but two instances, gains are 35--65\% and 32--56\% (small benchmarks), 12--32\% and 5--19\% (large benchmarks) when considering the cultivation feasible cases. 
The two cultivation infeasible cases (thus requiring a factory), qft-100 and heisenberg-225, improve 97--99\% on both metrics.

The weakest results, knn-129 and wstate-118 at $\varepsilon=10^{-3}$ (5\% and 11\% in space-time, 12\% and 14\% in cycles), are the two large instances whose per-\tstate{} budget cultivation can only just satisfy.
Their baseline already spends 50--57\% of the budget, so the fixed cutoff is close to what the budget requires in the first place.
The factories \name{} adds for throughput cost qubit counts that offset most of the cycle gain in the product.
Notably, however, there exist no cases in which \name{} is slower or larger than the baseline.

\subsection{Budget Expenditure}

\name{} spends essentially the entire budget in every run and never exceeds it.
In the bottom rows of the benchmark panels, we track the budget expenditure.
The blue and green bars, showing the expenditure for the minimum cycle count and minimum volume targets respectively, sit on the dashed line in all 64 configurations, the intended behavior of the dynamic \tstate{} post-selection we propose.
The baselines leave most of the budget unused, and how much they leave explains the size of \name{}'s advantage.

At a fixed $G=104$, the cultivation baseline spends 5--25\% of the budget on the small instances and 36--57\% on the large ones, rising as $\varepsilon/n_T$ falls.
\name{}'s improvement falls in step, from 65\% in cycles for ising-34 (baseline spends 5\%) to 35\% for wstate-76 (25\%), and to 12--32\% on the large instances: unspent budget is exactly the resource \name{} converts into speed.

The single-factory baseline spends far less of the budget. 
1--5\% on the small instances at $\varepsilon=10^{-5}$, 7--49\% on the large instances at $\varepsilon=10^{-5}$, and under 1\% for qft-100 and heisenberg-225 at $\varepsilon=10^{-3}$, because the smallest factory that meets the per-\tstate{} requirement typically overshoots it by an order of magnitude or more.
Here \name{}'s improvement stays at 88--95\% in cycles even as the baseline's expenditure grows, because a single factory is throughput-bound rather than budget-bound.
That is, the factory output rate, not fidelity, sets the cycle count.
\name{} removes the bottleneck with several factories plus in-channel cultivation at the cutoff the surplus budget permits, and the cycle count falls to the layout's congestion floor.

\section{Conclusion}
\label{sec:conclusion}

Allocating resources for \tstate{}-preparation requires solving three decisions between the two leading \tstate{} cultivation and distillation. 
How hard to post-select cultivation, which distillation factories to build, and how to schedule consumption against a total $T$-error budget.
We propose \name{} to resolve these decisions jointly.
The complementary gap prices each cultivated state, a size and quality characterizing catalog (extended by cultivating factories) replaces the $35p^3$ black box associated with \fone{} factories, and a fast heuristic allocator searches the joint space.
Across sixteen benchmarks and two budgets, \name{} beats both baselines in all 32 pairs, 69\% fewer logical cycles and 58\% less space-time volume on average (geometric-mean speedups of 5.5$\times$ in logical cycles and reductions of 3.4$\times$ in space-time volume), never exceeding the budget.

\begin{acks}
This work was supported by the U.S. Department of Energy (DOE) under Contract No. DE-AC02-05CH11231, through the Office of Advanced Scientific Computing Research EXPRESS Program grant FP00018751.
\end{acks}

\bibliographystyle{ACM-Reference-Format}
\bibliography{refs}

\appendix{}

\section{The Exact Allocation Dynamic Program}
\label{app:exact-dp}

This appendix lists the procedures of the exact allocation dynamic program of
Section~\ref{sec:allocation}.
The state, the three dominance statistics $(q, r, \varphi)$, and the validity
rules for configurations are defined in Section~\ref{sec:allocation}; the
algorithms here operate on them unchanged.

Algorithm~\ref{alg:plugins-op} gives the shared machinery: the dominance test,
the antichain filter \textproc{Nondom}, and the \textproc{ParetoMerge} and
\textproc{ParetoExtend} plug-ins that the scan invokes.
Algorithm~\ref{alg:edgescan-op} scans the candidate placements of one edge in
position order, maintaining one table per overhang key $\kappa = (c_L, c_R)$;
the take branch extends every configuration compatible with candidate $w_i$
(its predecessor $\pred(i)$ is the last candidate whose interval does not
conflict with $w_i$), and the skip branch carries configurations that omit it.
Algorithm~\ref{alg:ringjoin-op} joins the four per-edge tables around the
layout ring, discarding combinations whose corner flags conflict and applying
the dominance filter to the survivors.
Algorithm~\ref{alg:exactdp-op} is the top-level driver: it runs the edge
scan once per distinct edge length (rectangular layouts mirror the
remaining edges), joins the four tables around the ring, and assigns each
surviving vector its operating point by taking the lowest gap cutoff on
the precomputed ladder whose cultivation mixture fits the budget.
The result is the exact Pareto front over $(q, r, \varphi)$, each point
carrying its gap cutoff and a witness placement set recoverable by
following parent references.

\begin{algorithm}[h]
\caption{\textsc{ExactDP}}
\label{alg:exactdp-op}
\begin{algorithmic}[1]
\Require per-edge candidate placements from the catalog; budget
$\varepsilon$; $T$ count $n_T$; cutoff ladder $r_C(G), \bar{p}_C(G)$
\For{$e \in \{\mathrm{T}, \mathrm{R}, \mathrm{B}, \mathrm{L}\}$}
  \State $g_e \gets \Call{EdgeScan}{\text{candidates}(e)}$
  \Comment{rectangular layouts: compute one edge per distinct length,
  mirror the rest}
\EndFor
\State $Z \gets \Call{RingJoin}{g_{\mathrm{T}}, g_{\mathrm{R}},
  g_{\mathrm{B}}, g_{\mathrm{L}}}$
\For{$v \in Z$}
  \State $G^\ast(v) \gets \min\{G : (r_C(G)\bar{p}_C(G) + v.\varphi)
    / (r_C(G) + v.r) \le \varepsilon / n_T\}$
\EndFor
\State \Return $\{(v, G^\ast(v)) : v \in Z,\ G^\ast(v)$ exists$\}$,
  witnesses via parent back-pointers
\end{algorithmic}
\end{algorithm}

\begin{algorithm}[h]
\caption{Auxiliary Procedures}
\label{alg:plugins-op}
\begin{algorithmic}[1]
\Procedure{Dominates}{$v$, $u$}
  \Comment{$v, u$ are value records $(q, r, \varphi)$; assumes $v \ne u$}
  \State \Return $q_v \le q_u$ \textbf{and} $r_v \ge r_u$ \textbf{and}
    $\varphi_v \le \varphi_u$
\EndProcedure
\Procedure{Nondom}{$\mathit{pool}$}
  \State sort $\mathit{pool}$ by the stable key $(q$ ascending, $-r$,
    $\varphi)$; collapse exact-duplicate vectors, keeping the first
  \State \Return $\{ u \in \mathit{pool} :$ no $v \in \mathit{pool}$ has
    \textproc{Dominates}($v$, $u$)$\}$
\EndProcedure
\Statex
\Procedure{ParetoMerge}{$a$, $b$}
  \Comment{$a, b$: antichains of records}
  \State \Return \textproc{Nondom}(the concatenation of $a$ and $b$)
\EndProcedure
\Procedure{ParetoExtend}{$a$, $w$}
  \Comment{$w$: placement id, with costs $(q_w, r_w, \varphi_w)$}
\State \Return \textproc{Nondom}($[$\textproc{Record}($q_u + q_w$,
    $r_u + r_w$, $\varphi_u + \varphi_w$, $\mathit{parent} = u$,
    $\mathit{placement} = w$) for $u$ in $a]$)
\EndProcedure
\Statex

\end{algorithmic}
\end{algorithm}

\begin{algorithm}[h]
\caption{\textsc{EdgeScan}}
\label{alg:edgescan-op}
\begin{algorithmic}[1]
\Require candidate placements $w_1 \dots w_m$ of one edge in position order;
  predecessor map $\pred(i)$ (last candidate whose interval does not conflict
  with $w_i$); overhang flags $\pi_L(i), \pi_R(i)$; plug-in
  \textsc{Merge}, \textsc{Extend}, \texttt{MERGE\_ID}, \texttt{EXTEND\_ID}
  (Algorithm~\ref{alg:plugins-op})
\Ensure per-overhang-key tables $g_e[\kappa]$ of non-dominated valid
  configurations of the edge
\State $f_0[(0,0)] \gets \texttt{EXTEND\_ID}$;\ \
  $f_0[\kappa] \gets \texttt{MERGE\_ID}$ for $\kappa \ne (0,0)$
\For{$i \gets 1$ \textbf{to} $m$}
  \State $T[\kappa] \gets \texttt{MERGE\_ID}$ for all $\kappa$
    \Comment{take branch}
  \ForAll{$\kappa = (c_L, c_R) \in \{0,1\}^2$}
    \State $\kappa' \gets (c_L \lor \pi_L(i),\ c_R \lor \pi_R(i))$
    \State $T[\kappa'] \gets{}$ \textproc{Merge}($T[\kappa']$,
      \textproc{Extend}($f_{\pred(i)}[\kappa]$, $w_i$))
  \EndFor
  \ForAll{$\kappa \in \{0,1\}^2$}
    \State $f_i[\kappa] \gets{}$ \textproc{Merge}($f_{i-1}[\kappa]$, $T[\kappa]$)
      \Comment{skip branch merged in}
  \EndFor
\EndFor
\State \Return $g_e \gets f_m$
\end{algorithmic}
\end{algorithm}

\begin{algorithm}[h]
\caption{\textsc{RingJoin}}
\label{alg:ringjoin-op}
\begin{algorithmic}[1]
\Require edge tables $g_{\mathrm{T}}, g_{\mathrm{R}}, g_{\mathrm{B}},
  g_{\mathrm{L}}$ from Algorithm~\ref{alg:edgescan-op}; \textsc{Combine}
  (counting: convolution; Pareto: pairwise Minkowski sum then $\nondom$;
  slack: $+$); gate predicate $\mathrm{gate}(a, b) = \lnot (a \land b)$
\Ensure the non-dominated set over $(q, r, \varphi)$ of valid ring
  configurations, corner conflicts excluded

\Procedure{ParetoCombine}{$a$, $b$}
  \Comment{all-pairs sums across two antichains (Minkowski sum), then filter}
  \State \Return \textproc{Nondom}($[\,$\textproc{Record}($q_u + q_v$,
    $r_u + r_v$, $\varphi_u + \varphi_v$, $\mathit{parents} = (u, v)$)
    for $u$ in $a$, $v$ in $b \,]$)
  \Comment{joined records carry two parent refs;
    \textproc{ReconstructConfig} recurses into both}
\EndProcedure

\State $Z \gets \texttt{MERGE\_ID}$
\ForAll{$(t_L, t_R), (r_L, r_R), (b_L, b_R), (l_L, l_R) \in \{0,1\}^2$}
  \If{$\mathrm{gate}(t_R, r_L) \land \mathrm{gate}(r_R, b_L) \land
      \mathrm{gate}(b_R, l_L) \land \mathrm{gate}(l_R, t_L)$}
    \State $P \gets{}$ \textproc{Combine}(\textproc{Combine}($g_{\mathrm{T}}[t_L, t_R]$,
      $g_{\mathrm{R}}[r_L, r_R]$),
      \textproc{Combine}($g_{\mathrm{B}}[b_L, b_R]$, $g_{\mathrm{L}}[l_L, l_R]$))
    \State $Z \gets{}$ \textproc{Merge}($Z$, $P$)
  \EndIf
\EndFor
\State \Return $Z$
\end{algorithmic}
\end{algorithm}

\section{Placement Sensitivity}
\label{app:placement}

\begin{figure*}[t]
    \centering
    \includegraphics[width=\linewidth]{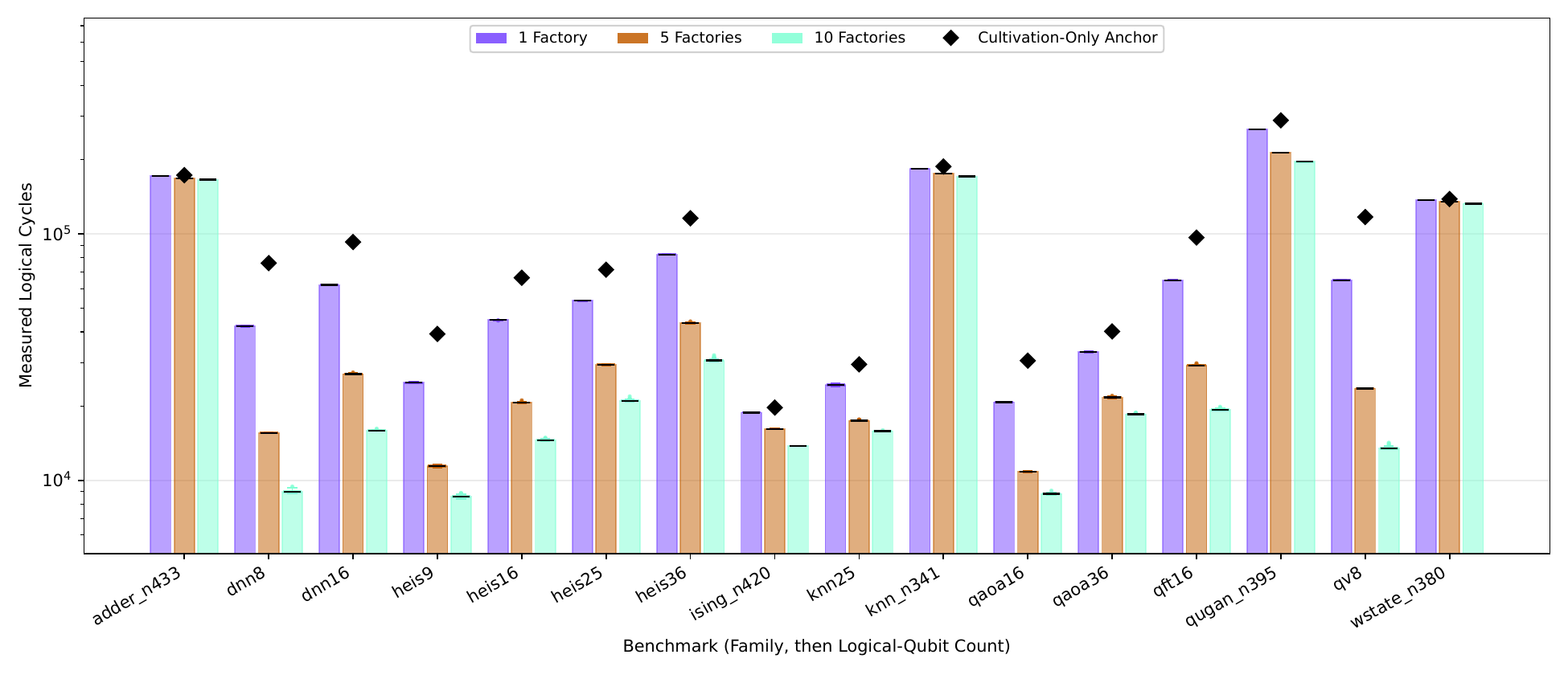}
    \Description{}
    \caption{Placement sensitivity of perfect-factory fleets under
    tandem cultivation with a dynamic budget ($\varepsilon = 10^{-3}$).
    Bars give the median measured logical cycles over sampled perimeter-port
    placements at fleet sizes 1, 5, and 10 (cadence $k = 3$); the box on each
    bar is the full placement distribution, and the black diamond is the
    cultivation-only anchor.
    Boxes are single-seed, so their spreads include cultivation
    stochasticity as well as placement; even so they are negligible against
    the fleet-size effect on every benchmark.}
    \label{fig:placement_bars}
\end{figure*}

Figure~\ref{fig:placement_bars} shows the placement-sensitivity sweep that
justifies dropping the placement dimension from the relaxed allocator of
Section~\ref{sec:method}.
Perfect factories at a fixed cadence of one \tstate{} every $3$ logical cycles
are seated on randomly sampled perimeter-port subsets of each layout, at fleet
sizes of 1, 5, and 10, with cultivation running in the routing channels under
the standard dynamic budget bundle at $\varepsilon = 10^{-3}$.
Across all sixteen programs, from 8 to 433 logical qubits, the spread across
placements (the box on each bar) is negligible against the effect of fleet
size, and against the cultivation-only anchor: \emph{where} factories sit on
the perimeter matters far less than \emph{how many} there are.
The largest programs also show fleet size itself saturating --- adder-433,
knn-341, and wstate-380 gain little from 10 factories over 1 --- the
congestion-bound regime discussed in Section~\ref{sec:results}.
\clearpage

\end{document}